\documentclass[10pt]{article}
\pdfoutput=1
\usepackage[affil-it]{authblk}
\usepackage{fullpage}
\usepackage{url}
\usepackage{amsmath}
\usepackage{amsfonts}
\usepackage{graphicx}
\usepackage{subcaption}
\usepackage{algorithm}
\usepackage{algorithmic}
\usepackage{hyperref}
\usepackage[percent]{overpic}
\usepackage{multirow}
\usepackage{xcolor}
\usepackage[UKenglish]{babel}
\DeclareMathOperator*{\argmin}{\arg\!\min}

\usepackage{natbib}
\newcommand{\BREAK}{\STATE \algorithmicbreak}
\newcommand{\algorithmicbreak}{\textbf{break}}
\begin{document}
	
	\title{Blur resolved OCT: full-range interferometric synthetic aperture microscopy through dispersion encoding}
	
	\author{Jonathan H. Mason\textsuperscript{1}%
	\thanks{Electronic address: \texttt{j.mason@ed.ac.uk}; Corresponding author} }
	\author{Mike E. Davies\textsuperscript{2}}\author{Pierre O. Bagnaninchi\textsuperscript{1}}
	\affil{\textsuperscript{1}MRC Centre for Regenerative Medicine, The University of Edinburgh, 5 Little France Drive, Edinburgh, EH16 4UU, UK}
	\affil{\textsuperscript{2}School of Engineering, Institute for Digital Communications, The University of Edinburgh, Edinburgh, EH9 3JL, UK}
	
	\maketitle
	
	
	\begin{abstract}
		We present a computational method for full-range interferometric synthetic aperture microscopy (ISAM) under dispersion encoding. With this, one can effectively double the depth range of optical coherence tomography (OCT), whilst dramatically enhancing the spatial resolution away from the focal plane. To this end, we propose a model-based iterative reconstruction (MBIR) method, where ISAM is directly considered in an optimization approach, and we make the discovery that sparsity promoting regularization effectively recovers the full-range signal. Within this work, we adopt an optimal nonuniform discrete fast Fourier transform (NUFFT) implementation of ISAM, which is both fast and numerically stable throughout iterations. We validate our method with several complex samples, scanned with a commercial SD-OCT system with no hardware modification. With this, we both demonstrate full-range ISAM imaging, and significantly outperform combinations of existing methods.
	\end{abstract}
	
	\section{Introduction}
	Optical coherence tomography (OCT) offers high resolution non-invasive imaging of tissues and other semi-transparent materials \cite{Huang1991,Fujimoto2000,Tomlins2005,Liu2017}. Through the reflection interference between a reference and sample arm, the structure of scatterers along depth are encoded. In spectral-domain OCT (SD-OCT) \cite{Wojtkowski2004}, this interferometry signal is diffracted onto a detector array, from which the one-dimensional structure (A-scan) can be reconstructed through an inverse fast discrete Fourier transform (IFFT). The three-dimensional structure of the specimen can then be formed by raster scanning the sample and combining the resulting profiles.
	
	One deficit of this simplistic scheme is that A-scans are not independent, due to the widening of the beam away from the focal plane of the lens in the sample arm. With this, objects appear blurred in the out-of-focus region of the image, leading to a non-uniform resolution with depth. With interferometric synthetic aperture microscopy (ISAM), Ralston et al. \cite{Ralston2006,Ralston2006b,Ralston2007} showed that this effect may be actively compensated for by resampling the spatial frequency components and filtering, which effectively combines information from neighboring A-scans, and leads to a uniform resolution with depth.
	
	Another potential deficit of SD-OCT is due to the measurements at the spectrometer being real intensities. Therefore, its Fourier transform will be conjugate symmetric, effectively halving the available depth range. In practice, one often ensures the absence of objects in the negative optical delay region, and then ignore the superfluous mirror image after applying an IFFT. There are several hardware approaches to utilize the entire range, such as placing a phase modulator in the sample arm \cite{Gotzinger2005,Kim2010}, offsetting the scanning mirror pivot \cite{An2007}, or measuring the quadrature component of the interferometry signal \cite{Mao2008}, although these solutions increase system complexity and reduce scanning rate due to requiring several measurements \cite{Hofer2009}.
	
	It is also possible to differentiate the conjugate terms, by introducing a dispersion discrepancy between sample and reference arms. This is well approximated as a non-linear phase lag, which acts in an opposite direction for the mirrored complement. Therefore, after compensating for dispersion in one direction, the mirror component becomes `doubly dispersed' leading to a blurring and distinction from the desired sharpened signal. In dispersion encoded full-range (DEFR) OCT \cite{Hofer2009}, one takes a greedy optimization approach to resolving the object, by iteratively removing the blurred mirror associated with the highest magnitude component. This uses the implicit approximation that A-scans are independent. There have been several extensions to this, including removing several components on each iteration \cite{Witte2009,Hofer2010}, and removing autocorrelation artefacts also \cite{Kottig2012}. It was recently shown that DEFR may indeed allow faithful reconstruction even under subsampling regimes \cite{Yi2018}. Interestingly, there are strong parallels between these approaches and radio frequency interference suppression in synthetic aperture radar (SAR) \cite{Kelly2013}.
	
	In order to perform full-range imaging from real spectral measurements, one must accept an inherent sampling deficit from inferring as many complex parameters as real samples, which is the case in DEFR methods. As commonly employed in the field of compressed sensing \cite{Candes2006a,Donoho2006b}, one can introduce a sparsity constraint that allows the faithful reconstruction of sparse signals from few measurements. It has been demonstrated that images from OCT are typically sparse in some domain, and compressed sensing restoration methods have been proven successful \cite{Hofer2009,Liu2010,Mohan2010,Mason2019}. 
	
	It was recently shown in \cite{Mason2019} that the ISAM resampling can be utilized in a model-based iterative reconstruction (MBIR) in the half-range setting under sub-sampling, with sparsity promoting regularization as is used in compressed sensing. In this work, we extend MBIR to the dispersion encoded full-range setting. We present an accelerated MBIR algorithm, along with an enhanced variation, and evaluate it with synthetic and real full-range measurements of several complex samples.
	
	\subsection{Novel contributions}
	We demonstrate, for the first time, full-range dispersion encoded ISAM. As well as showing how this may be achieved through a naive combination of two existing algorithms (DEFR+ISAM), we propose a novel MBIR optimization approach as a solution. Unlike the greedy DEFR methods, this has the potential to exploit the shared information between A-scans, such as multidimensional sparsity in the ISAM refocussed space. We provide analysis from quantitative simulation and real data, where we show the feasibility of reconstructing refocussed full-range measurements with computational methods, and observe a significant performance gain of MBIR over alternative approaches.
	
	\section{Background}
	In this section, we describe the dispersion encoded full-range ISAM model exploited by our MBIR method. Here, we wish to reconstruct the complex susceptibility, $\boldsymbol{\eta}\in\mathbb{C}^{N}$, from the real spectrometer measurements, $\boldsymbol{s}\in\mathbb{R}^{N}$, where $N = N_xN_yN_z$ with $N_x$ and $N_y$ the number of lateral measurements in $x$ and $y$, and $N_z$ the number of axial measurements (also the resolution of the spectrometer). Since we wish to infer $N$ complex numbers with $2N$ unknowns from only $N$ measurements, this represents a clear sampling deficit, which we attempt to overcome by exploiting sparsity.
	
	The ISAM model tells us that the 3D Fourier transform of the object, given as
	\begin{equation}
	{\boldsymbol{H}}(\boldsymbol{q}_x,\boldsymbol{q}_y,\boldsymbol{\beta}) = \mathcal{F}_\mathrm{3D}(\boldsymbol{\eta}(\boldsymbol{x},\boldsymbol{y},\boldsymbol{z})),
	\end{equation}
	where $\boldsymbol{q}_x$, $\boldsymbol{q}_y$ and $\boldsymbol{\beta}$ are transverse and axial spatial frequencies respectively, is related to the transverse Fourier transform of the complex interferometry signal
	\begin{equation}
	\boldsymbol{S}(\boldsymbol{q}_x,\boldsymbol{q}_y,\boldsymbol{k}) = \mathcal{F}_\leftrightarrow(\boldsymbol{s}_c),
	\end{equation}
	where $\boldsymbol{s}_c$ is the complex interferometry signal, of which we directly measure its real part; we use the notation $\mathcal{F}_\leftrightarrow(\cdot)$ for the Fourier transform in the transverse spatial dimensions, and $\boldsymbol{k}$ is the wavenumber.
	
	The ISAM relationship can then be expressed as \cite{Ralston2008}
	\begin{equation} \label{equ:isam_ft}
	\boldsymbol{S}(\boldsymbol{q}_x,\boldsymbol{q}_y,\boldsymbol{k}) = \boldsymbol{B}(\boldsymbol{q}_x,\boldsymbol{q}_y,\boldsymbol{k})\odot{\boldsymbol{H}}\left(\boldsymbol{q}_x,\boldsymbol{q}_y,\boldsymbol{\beta}\right)|_{\boldsymbol{\beta} = -\sqrt{4k^2-\boldsymbol{q}_x^2-\boldsymbol{q}_y^2}},
	\end{equation}
	where $\boldsymbol{B}(\boldsymbol{q}_x,\boldsymbol{q}_y,k)$ is a filter to account for the lens and intensity drop off \cite{Davis:07b} and a frequency warping effect is seen through the relation $\beta = -\sqrt{4k^2-\boldsymbol{q}_x^2-\boldsymbol{q}_y^2}$, $\odot$ represents element-wise multiplication. This resampling is known as the Stolt mapping, and is used in the fields of seismology and SAR \cite{Stolt1978,Davis2008}.
	
	Since the resampling through interpolation, filtering and Fourier transform are all linear operations, we can express this by \cite{Hofer2009}
	\begin{equation} \label{equ:isam}
	\boldsymbol{s}_c = \boldsymbol{K}\boldsymbol{\eta},
	\end{equation}
	where $\boldsymbol{K}$ is a matrix representing the mapping from susceptibility image to the complex spectroscopic signal.
	
	We only directly measure the real part of this signal \cite{Tomlins2005}, which we can express as
	\begin{equation}
	\boldsymbol{s} = \Re(\boldsymbol{K}\boldsymbol{\eta}),
	\end{equation}
	where $\Re(\cdot)$ selects the real part. This is equivalent to
	\begin{equation} \label{equ:mirror_conj}
	\boldsymbol{s} = \frac{1}{2}\left[\boldsymbol{s}_c+\bar{\boldsymbol{s}}_c^*\right],
	\end{equation}
	where $\bar{\boldsymbol{s}}_c^*$ is the complex spectrum from the conjugate component we wish to suppress. In practice, there will also be background and autocorrelation signals on top of $s$ measured at the spectrometer. Throughout this work, we assume the background can be easily removed and the autocorrelation components are small. Since autocorrelation artifacts have been shown to be influenced in both ISAM \cite{Davis:07a} and DEFR \cite{Kottig2012}, their treatment may be useful in highly scattering specimens.
	
	In either the half-range of full-range setting \cite{Wojtkowski2004,Hofer2009}, when a dispersion discrepancy between sample and reference arms exists, this may be accurately modelled through a non-linear phase term given as
	\begin{equation} \label{equ:disp_poly}
	e^{j\boldsymbol{\phi}} = \exp\left(j\sum_{i=1}^{N_p} a_i(\boldsymbol{k}-k_0)^i\right),
	\end{equation}
	where $k_0$ is the central frequency component, $a_i$ are the polynomial coefficients, and $N_p$ is the order. In practice, $N_p = 3$ is usually sufficient to capture significant dispersion effects and $\boldsymbol{a}$ may be found through an autofocus algorithm \cite{Wojtkowski2004}. With this, Eq.~(\ref{equ:mirror_conj}) becomes
	\begin{equation} \label{equ:disp_relation}
	\boldsymbol{s}_d = \frac{1}{2}\left[\boldsymbol{s}_c\odot e^{j\boldsymbol{\phi}}+\bar{\boldsymbol{s}}_c^*\odot e^{-j\boldsymbol{\phi}}\right],
	\end{equation}
	where $\boldsymbol{s}_d$ represents the real measurements as in Eq.~(\ref{equ:mirror_conj}) with the inclusion of dispersion, and the phase shift has opposite effect on each of the conjugate components \cite{Hofer2009}. We highlight that the derivation of Eq.~(\ref{equ:disp_relation}) as in \cite{Hofer2009} uses a thin sample approximation, whereby dispersion between the layers of the sample is not considered.
	
	In standard half-range imaging, dispersion compensated reconstruction can be performed by
	\begin{equation} \label{equ:double_disperse}
	\mathcal{F}^{-1}_\updownarrow(\boldsymbol{s}_d\odot e^{-j\boldsymbol{\phi}}) = \frac{1}{2}\left[\mathcal{F}^{-1}_\updownarrow(\boldsymbol{s}_c)+\mathcal{F}^{-1}_\updownarrow(\bar{\boldsymbol{s}}_c^*\odot e^{-2j\boldsymbol{\phi}})\right],
	\end{equation}
	where we use the notation $\mathcal{F}^{-1}_\updownarrow$ for an IFFT in the axial dimension. If the object of interest only occupies the positive delay area, then $\mathcal{F}^{-1}_\updownarrow(\bar{\boldsymbol{s}}_c^*\odot e^{-2j\boldsymbol{\phi}})$ will not overlap with the desired signal, and can be easily ignored. When there is an overlap, and given sufficient dispersion encoding, then DEFR \cite{Hofer2009} can approximately remove the unwanted blurred artefacts in an iterative manner.
	
	\subsection{DEFR+ISAM} \label{sec:defr_isam}
	One approach we offer to achieve full-range ISAM under dispersion encoding, is through combining the two existing algorithms as presented in \cite{Hofer2009,Ralston2008}.
	
	Firstly, DEFR attempts to solve the following optimization problem based on the dispersion encoding from Eq.~(\ref{equ:double_disperse}) in a greedy fashion
	\begin{equation} \label{equ:defr_obj}
	\hat{\boldsymbol{z}} = \argmin_{\boldsymbol{z}} \|2\Re\{\mathcal{F}_\updownarrow(\boldsymbol{z})\odot e^{j\boldsymbol{\phi}}\}-\boldsymbol{s}_d\|_2^2,
	\end{equation}
	The iterative method in \cite{Hofer2009} works like Matching Pursuit (MP) \cite{Mallat1993,Duarte2006}, by updating $\boldsymbol{z}$ one component at a time, giving a solution that is sparse.
	
	From here, one can then estimate the susceptibility by applying the ISAM back-projection operator from Eq.~(\ref{equ:isam}) as
	\begin{equation}
	\boldsymbol{\eta} = \boldsymbol{K}^H(\mathcal{F}_\updownarrow(\hat{\boldsymbol{z}})+\boldsymbol{r}),
	\end{equation}
	where $\boldsymbol{r}$ is the residual term from the approximate solution of Eq.~(\ref{equ:defr_obj}) as defined in \cite{Hofer2009}.
	
	Since DEFR operates on each A-scan independently, it is unable to exploit multidimensional sparsity in the image domain. This is especially relevant when applied before ISAM, as one expects significant portions of the image to be out of focus, which will decrease the observed sparsity. Secondly, the application of Eq.~(\ref{equ:isam}) calculates the back-projection, which is only a crude approximation of the inverse of the system model in $\boldsymbol{K}$.

	\section{Method}
	To overcome the potential deficits of the simple two-step approach suggested in Section~\ref{sec:defr_isam}, we propose that full-range DEFR--ISAM can be achieved by solving of the following optimization problem
	\begin{equation} \label{equ:mbir}
	\tilde{\boldsymbol{\eta}} = \argmin_{\boldsymbol{\eta}}\frac{1}{2}\|\Re(\hat{\boldsymbol{K}}\boldsymbol{\eta})-\boldsymbol{s}_d\|_2^2 + \lambda \boldsymbol{w}^T|\boldsymbol{\eta}|,
	\end{equation}
	where $\boldsymbol{w}^T|\boldsymbol{\eta}|$ is a weighted $\ell_1$ norm with the weighting vector $\boldsymbol{w}\in\mathbb{R}^N$. The unweighted $\ell_1$ penalty function --- usually denoted as $\|\boldsymbol{\eta}\|_1$ --- more commonly employed in compressed sensing can be achieved through choosing $\boldsymbol{w} = \boldsymbol{1}$, although choosing a nonuniform $\boldsymbol{w}$ can account for the increased sparsity with depth typical in OCT. We also adopt the compact notation in Eq.~(\ref{equ:mbir}) for the dispersion corrected ISAM model
	\begin{equation}
	\hat{\boldsymbol{K}} \equiv \mathrm{diag}(e^{j\boldsymbol{\phi}})\boldsymbol{K}.
	\end{equation}
	
	The $\ell_1$ penalty term in Eq.~(\ref{equ:mbir}) assumes the specimen is spatially sparse after refocussing and the conjugate artifacts have been removed. However, this method and algorithm are also easily extendable for any convex regularization function, such as total variation (TV) and wavelet sparsity \cite{Mason2019}, which may be more appropriate for specimens with different spatial structure.
	
	Problems with the form of Eq.~(\ref{equ:mbir}) have been extremely well studied in the field of compressed sensing \cite{Candes2006a,Donoho2006b}, in which many algorithms for its solution have been developed. In this work, we opt for the fast iterative thresholding shrinkage algorithm (FISTA) \cite{Beck2009}, which is an accelerated gradient descent method with soft-thresholding to minimize the $\ell_1$ function. FISTA applied to the objective function in Eq.~(\ref{equ:mbir}) is given in Algorithm~\ref{alg2}, with the weighted soft-thresholding operator defined as
	\begin{equation}
	\mathcal{T}_{\lambda w_i}(u_i) = \frac{u_i\max\left(|u_i|-\frac{\lambda w_i}{N},0\right)}{\max\left(|u_i|-\frac{\lambda w_i}{N},0\right)+\frac{\lambda w_i}{N}}.
	\end{equation}
	
	\begin{algorithm}                      
		\caption{MBIR full-range ISAM}
		\label{alg2}
		\begin{algorithmic}                    
			\REQUIRE Regularization constant $\lambda$, and starting point $\boldsymbol{\eta}^1 = \boldsymbol{z}^0 = \mathbf{0}$.
			\FOR{$k = 1,2,\ldots$}
			\STATE $\boldsymbol{z}^{k} \leftarrow \mathbf{\mathcal{T}}_{\lambda\boldsymbol{w}}\left(\boldsymbol{\eta}^{k}-\frac{1}{N}\hat{\boldsymbol{K}}^H\left(\Re(\hat{\boldsymbol{K}}\boldsymbol{\eta}^{k})-\boldsymbol{s}_d\right) \right)$ \COMMENT{gradient descent and thresholding step}
			\IF{$\|r_r^k\|<tol$ from Eq.~(\ref{equ:rel_res})}
			\BREAK
			\ENDIF
			\STATE $t^{k+1} \leftarrow \frac{1+\sqrt{1+4(t^k)^2}}{2}$
			\STATE $\boldsymbol{\eta}^{k+1} \leftarrow \boldsymbol{z}^{k}+\left(\frac{t^k-1}{t^{k+1}}\right)(\boldsymbol{z}^{k}-\boldsymbol{z}^{k-1})$ \COMMENT{update with momentum for convergence acceleration}
			
			\ENDFOR
			\STATE $\boldsymbol{\eta}\leftarrow \boldsymbol{\eta}^k+\hat{\boldsymbol{K}}^H\left(\boldsymbol{s}_d-\Re(\hat{\boldsymbol{K}}\boldsymbol{\eta}^{k})\right)$ \COMMENT{optional: add residual to solution}
			\RETURN $\boldsymbol{\eta}$
			
		\end{algorithmic}
	\end{algorithm}
	
	\subsection{Parameters}
	There are two elements in Algorithm~\ref{alg2} that must be appropriately chosen: the regularization constant $\lambda$, and the termination condition.
	
	For terminating the method we adopt the relative residual stopping condition from \cite{DBLP:journals/corr/GoldsteinSB14} defined through the value
	\begin{equation}\label{equ:rel_res}
	r_r^k = \frac{\|\boldsymbol{z}^k-\boldsymbol{\eta}^k\|}{\max\left\{\|\boldsymbol{g}^k\|,\|\boldsymbol{z}^k-\boldsymbol{\eta}^k+\boldsymbol{g}^k\|\right\}+\epsilon},
	\end{equation}
	where $\boldsymbol{g}^k = \frac{1}{N}\hat{\boldsymbol{K}}^H\left(\Re(\hat{\boldsymbol{K}}\boldsymbol{\eta}^{k})-\boldsymbol{s}\right)$, and $\epsilon$ is a small constant to ensure a non-zero denominator. With this, one terminates the iterations once $\|r_r^k\|<tol$, where $tol$ is the desired tolerance for convergence (we use $1\times 10^{-3}$ in our testing).
	
	
	%
	
	
	The regularization parameter $\lambda$, in combination with the optional weighting vector $\boldsymbol{w}$, will control the level of sparsity and hence conjugate artifact suppression in the reconstruction. It will also have an impact on the convergence, with larger values of $\lambda$ typically requiring fewer iterations to reach a given $tol$. From our testing on a range of different samples, we find $[0.1,0.8]$ to be a range providing good performance.
	
	The introduction of the weighting term $\boldsymbol{w}$ gives the option to use non-uniform regularization throughout the reconstructed image. From numerical and empirical testing, we have found an increase in $\boldsymbol{w}$ with depth produces better images. An intuition for this is that there are typically fewer photons collected from deep in the specimen due to scattering in the top layers, hence the local signal--to--noise ratio is reduced, so one should give less weight to the data fidelity term and more to the sparsity prior. We find that using a simple linearly increasing $\boldsymbol{w}$ with depth from 0.5 to 1, provides a large performance gain in our quantitative experiment presented in Section~\ref{sec:quant}. However, we suggest that other weighting schemes may be superior for different lenses or focal plane positions.
	
	As an optional final step in Alg.~\ref{alg2}, one can add the final residual onto the solution after termination of the MBIR. This is similar to the inclusion of residual in DEFR \cite{Hofer2009}, and is also used in radar imaging \cite{Kelly2012}. The advantage of this is to retain non-sparse low intensity but informative signal, which may otherwise be suppressed by the regularization, but is too small to contribute significant conjugate artifacts. For applications such as optical coherence elastography \cite{Kennedy2017}, retaining the low level speckle structure is critical for displacement tracking. Another advantage of including the residual term is that more aggressive regularization can be used within the MBIR optimization, which results in faster convergence and hence faster reconstruction speeds.
	
	\subsection{Efficient and robust ISAM through non-uniform FFT}
	The proposed approach requires repeated realization of the ISAM model within the optimization algorithm and therefore it is necessary to have an accurate yet efficient implementation of the ISAM operator. In this work, this is realized through the non-uniform FFT (NUFFT) algorithm \cite{Fessler2003a}.
	
	We rewrite the ISAM operator as
	\begin{equation}
	\boldsymbol{K}\cdot \equiv N_r\mathcal{F}_\leftrightarrow^{-1}(\mathrm{diag}(\boldsymbol{b})\mathcal{N}(\cdot)),
	\end{equation}
	with the matrix $\boldsymbol{K}$ as in Eq.~(\ref{equ:isam}), $\mathcal{N}(\cdot)$ is the NUFFT operator and $\boldsymbol{b}$ is vector representation of the filter $\boldsymbol{B}(\boldsymbol{q},\boldsymbol{k})$ in Eq.~(\ref{equ:isam_ft}). We will henceforth treat the unfiltered solution in this work, whereby we exclude $\boldsymbol{b}$, which has been shown to have minimal qualitative effect \cite{Ralston2006,Ralston2008}.
	
	For a standard ISAM as in \cite{Ralston2008}, this is equivalent to back-projection as
	\begin{equation} \label{equ:bp}
	\boldsymbol{K}^H\boldsymbol{s} = \mathcal{N}'(\mathcal{F}_\leftrightarrow(\boldsymbol{s})).
	\end{equation}

	\section{Experiments}
	\subsection{Materials and measurements} \label{sec:mat}
	All measurements were acquired using a Wasatch Photonics 800nm SD-OCT system, with 2048 spectrometer elements. The system's imaging arm included a length of optical fibre, sufficient to introduce a dispersion discrepancy against the reference mirror. We recorded 1024 A-scans over a 2 mm lateral distance using its standard protocol, and extracted the raw spectrometer data for processing. In each case the focal point was adjusted, by eye, to lie within the sample, and at the zero delay position.
	
	Preprocessing from the raw data consisted of background subtraction, obtained by averaging across A-scans, and non-linear calibration from detector element to wavenumber, according to parameters from the manufacturer.
	
	The samples used were as follows:
	\begin{enumerate}
		\item Beaded gel: TiO$_2$ micro-beads suspended in 2\% agarose gel, at a concentration of 1 mg/ml. The powdered TiO$_2$ ($<$5 $\mu$m,  Sigma-Aldrich) was dispersed throughout the gel before curing, through combination of stirring, pipette agitation and sonication.
		\item Tape: a roll of GiftWrap Scotch adhesive tape.
		\item Cucumber: a slice of cucumber flesh sectioned and blotted to remove excess moisture.
	\end{enumerate}
	
	The reconstruction was performed retrospectively on a commodity PC with an Intel i7-8700 CPU, 16 GB of RAM, and an NVIDIA Titan Xp GPU. All software was written in and run with Matlab. 
	
	\subsection{Methods under test} \label{sec:methods}
	The various methods and their implementation that we analyze in this experimental section are as follows:
	\begin{itemize}
		\item \textit{Direct IFFT}: dispersion compensation is applied to the measurements, with the polynomial coefficients $a_2$ and $a_3$ in Eq.~(\ref{equ:disp_poly}) found through the autofocus method of \cite{Wojtkowski2004}, followed by an IFFT. We then used the same dispersion parameters in each of the following methods.
		\item \textit{DEFR}: the method as described in \cite{Hofer2009}, with the residual included as it resulted in a superior quantitative performance. We ran the algorithm for 500 iterations to ensure convergence.
		\item \textit{ISAM}: we follow dispersion compensation with ISAM reconstruction as described in \cite{Ralston2008}, through the NUFFT adjoint operator in Eq.~(\ref{equ:bp}), although without the cropping step to reduce the data to half-range.
		\item \textit{DEFR+ISAM}: the combination of DEFR followed by full-range ISAM. Although this method is not described in the literature, it is a naive approach that we use as a point of comparison for our proposed MBIR.
		\item \textit{MBIR}: a simple implementation of Alg.~\ref{alg2}, with no weighting ($\boldsymbol{w} = \boldsymbol{1}$), and the residual step included. In every case we use $\lambda=0.5$, which provided a good balance between numerical accuracy and visual clarity, and use the termination condition $tol=1\times10^{-3}$.
		\item \textit{MBIR+}: a more advanced implementation of Alg.~\ref{alg2} with the weighting term $\boldsymbol{w}$ linearly increasing with depth from 0.5 to 1, and the residual term also included. We find the residual term allows for more aggressive regularization and hence used $\lambda=0.5$ throughout the simulated and real tests, although this resulted in visible suppression of structure in direct MBIR.
	\end{itemize}
	
	\subsection{Quantitative test} \label{sec:quant}
	\begin{figure}[!htb]
		\centering
		\includegraphics[width=1\textwidth]{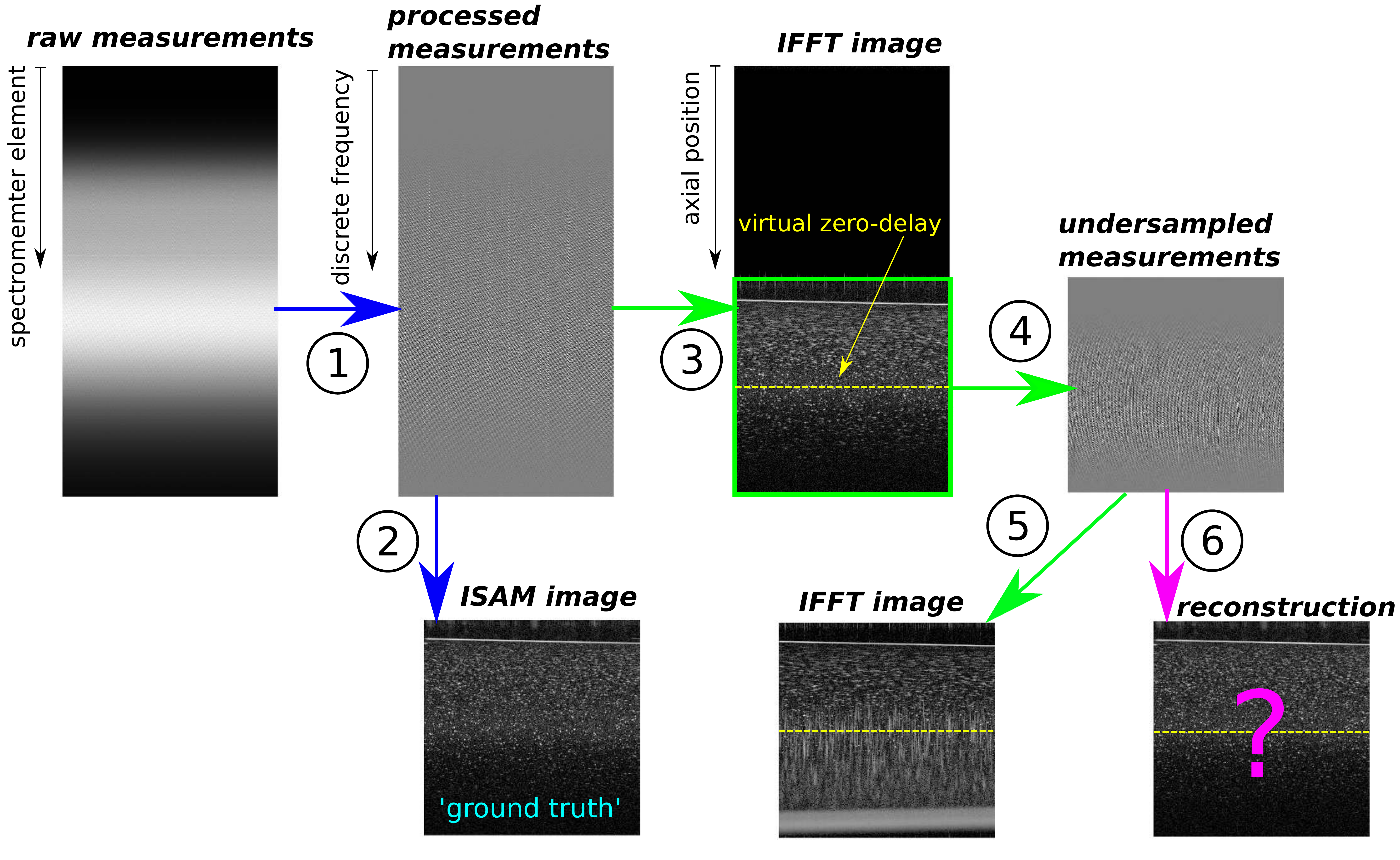}
		\caption{\label{fig:flow}Flow diagram showing the generation of synthetic full-range data from real measurements with ground truth. Descriptions of each processing step are detailed in the enumerated list in Sec.~\ref{sec:quant}.}
	\end{figure}
	
	To objectively assess the performance of our approach relative to the alternatives described in Sec.~\ref{sec:methods}, we performed analysis from pseudo-full-range data, against a ground truth obtained from real half-range measurements. In generating these measurements, we adjusted the focal plane to approximately lie halfway through the positive delay range, and ensured the negative delay was absent of scattering media. This process is illustrated in Fig~\ref{fig:flow}, with each step detailed as follows: 
	
	\begin{enumerate}
		\item \textbf{Preprocessing}: the raw measurements are processed to remove background, resample from detector element to discrete frequency (k-space), a Hilbert transform is taken to ensure zero negative delay, and finally dispersion compensation is applied.
		\item \textbf{ISAM mapping}: the processed measurements are reconstructed through fully sampled ISAM. This image represents a ground truth against which subsampled reconstructions can be assessed. The ground truth is also scaled by a factor 0.5, to match the magnitude change from taking real part during measurement synthesis.
		\item \textbf{IFFT}: an IFFT is applied to the processed data to map into image space. Due to the Hilbert transform, all values in the negative delay region will be zero.
		\item \textbf{Measurement synthesis}: the negative delay region of the IFFT image is truncated to place a virtual zero-delay at the dashed yellow line. The pseudo-full-range measurements are then generated by taking the axial FFT, applying the dispersion factor, and finally taking the real part.
		\item \textbf{Direct mapping}: directly applying an IFFT to the under-sampled measurements produces conjugate artifacts and lateral blurring in out--of--focus regions. We include this to demonstrate the magnitude of the artifacts, and the relative performance of the computational approaches.
		\item \textbf{Reconstruction}: we produce reconstructions with our proposed and tested methods described in Sec.~\ref{sec:methods} and evaluate their quantitative accuracy against the ground truth.  
	\end{enumerate}
	
	\begin{figure}[!htbp]
		\centering
		\begin{subfigure}[b]{0.32\textwidth}
			\begin{overpic}[width=\textwidth]{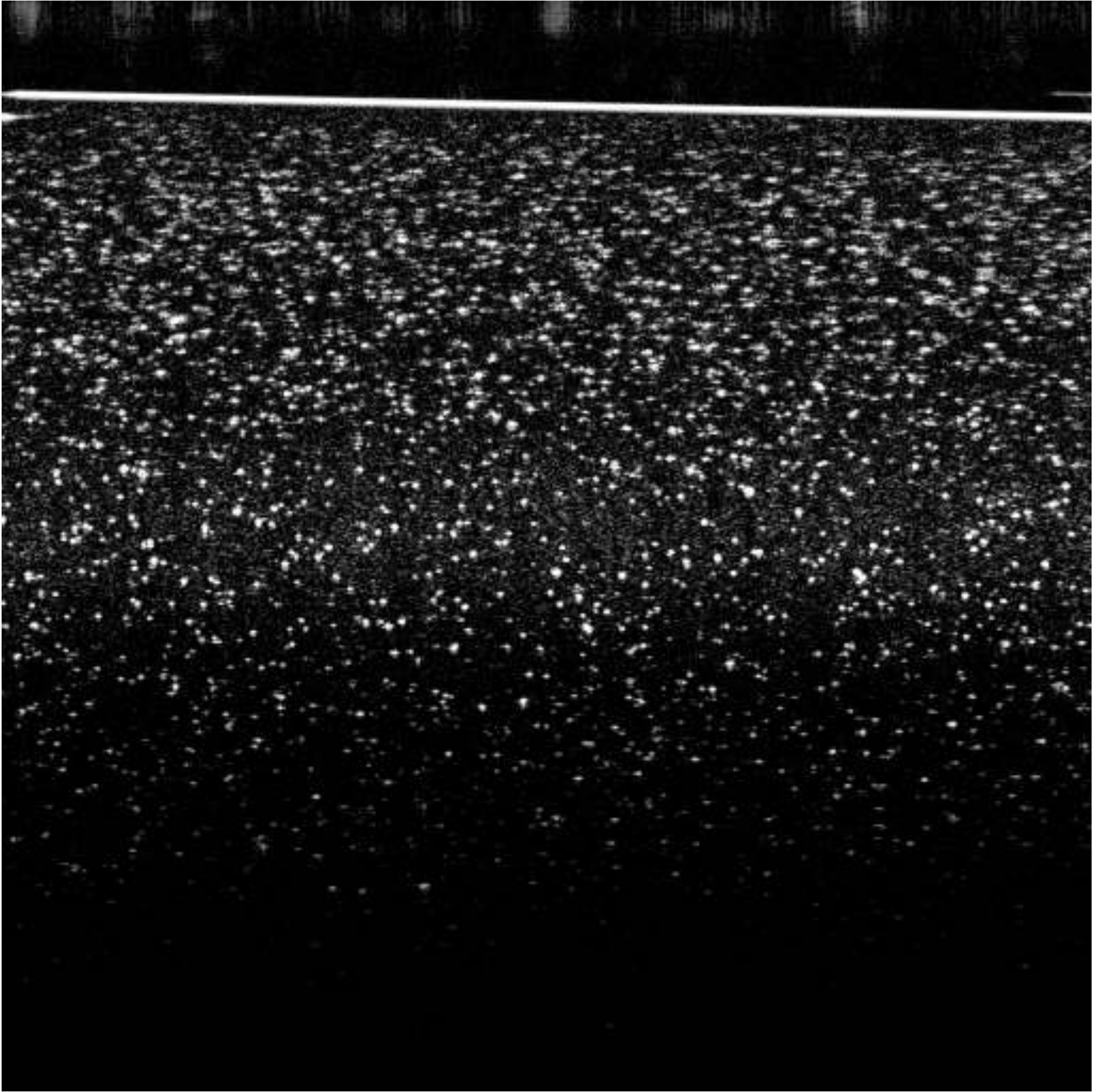}
				\includegraphics[width=\textwidth]{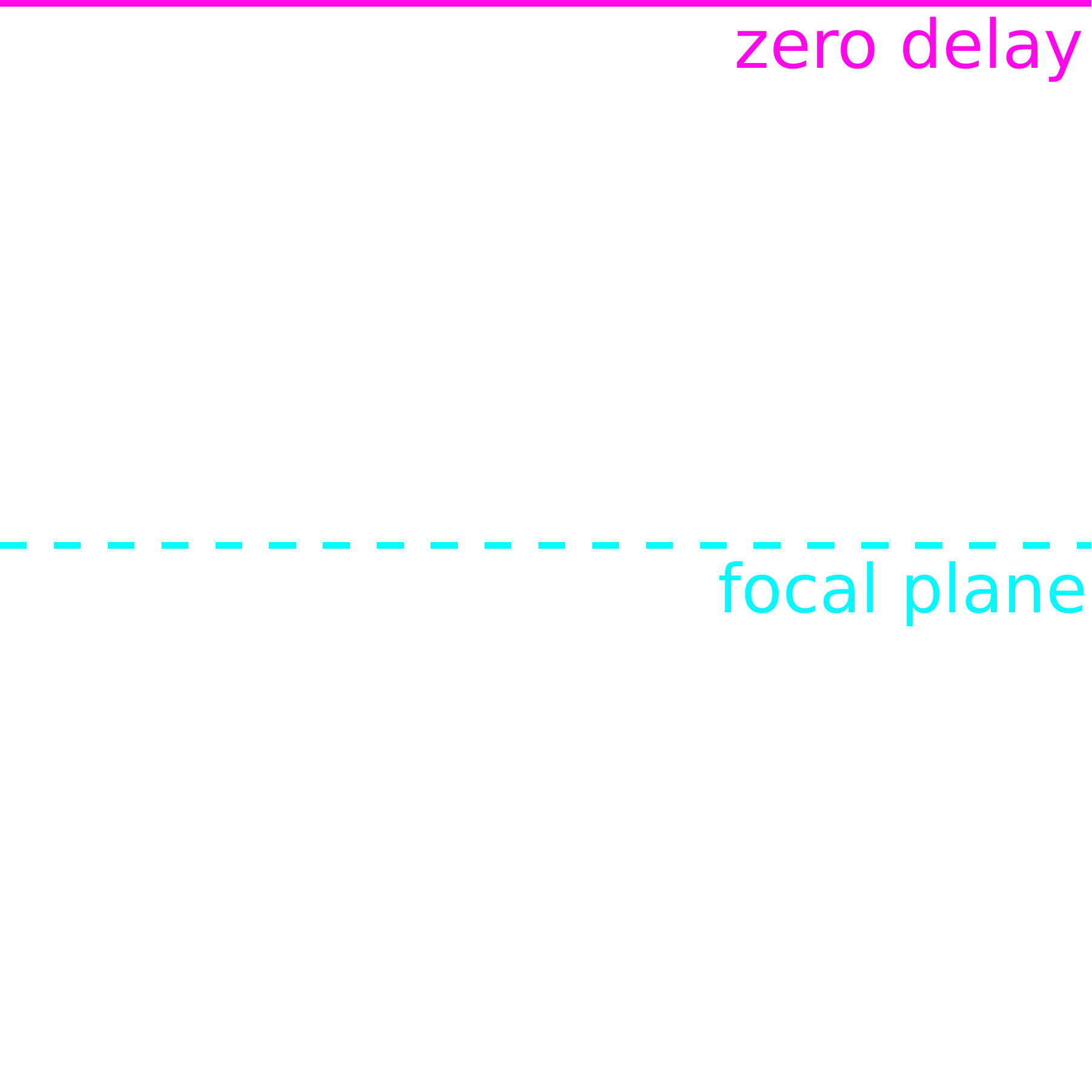} 
			\end{overpic}

			\caption{ground truth}
		\end{subfigure}
		\begin{subfigure}[b]{0.32\textwidth}
			\begin{overpic}[width=\textwidth]{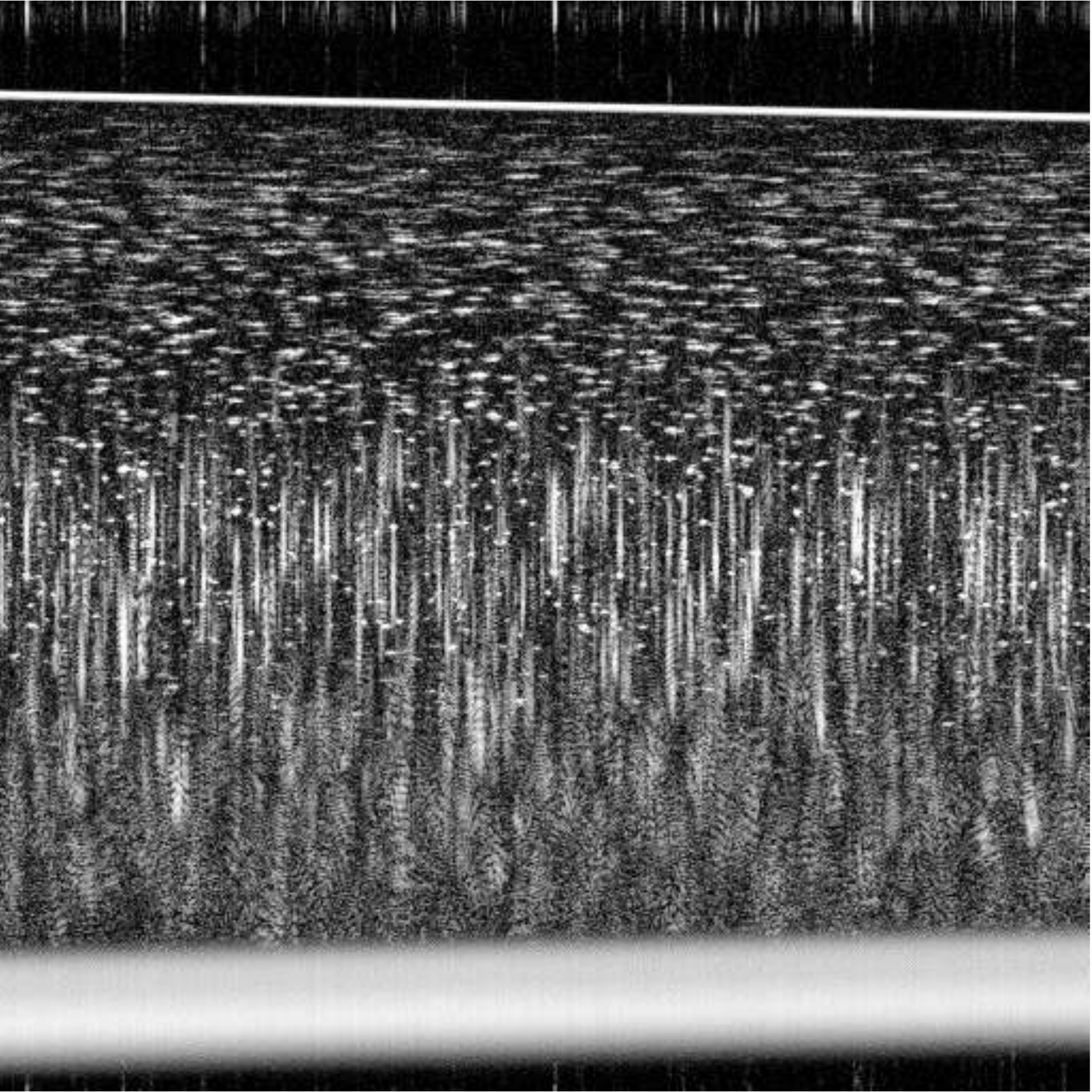}
				\includegraphics[width=\textwidth]{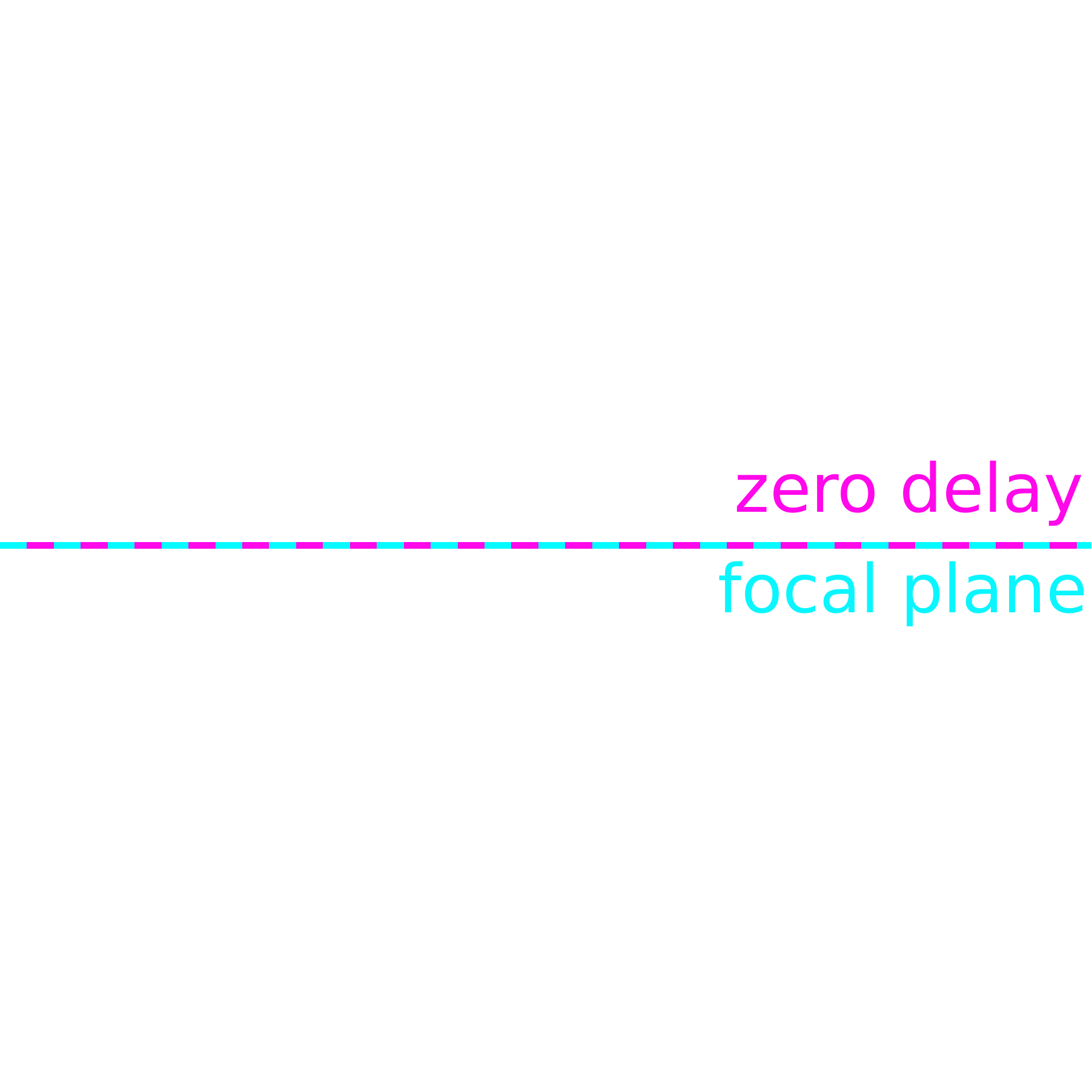} 
			\end{overpic}
			\caption{direct IFFT}
		\end{subfigure}
		\begin{subfigure}[b]{0.32\textwidth}
			\begin{overpic}[width=\textwidth]{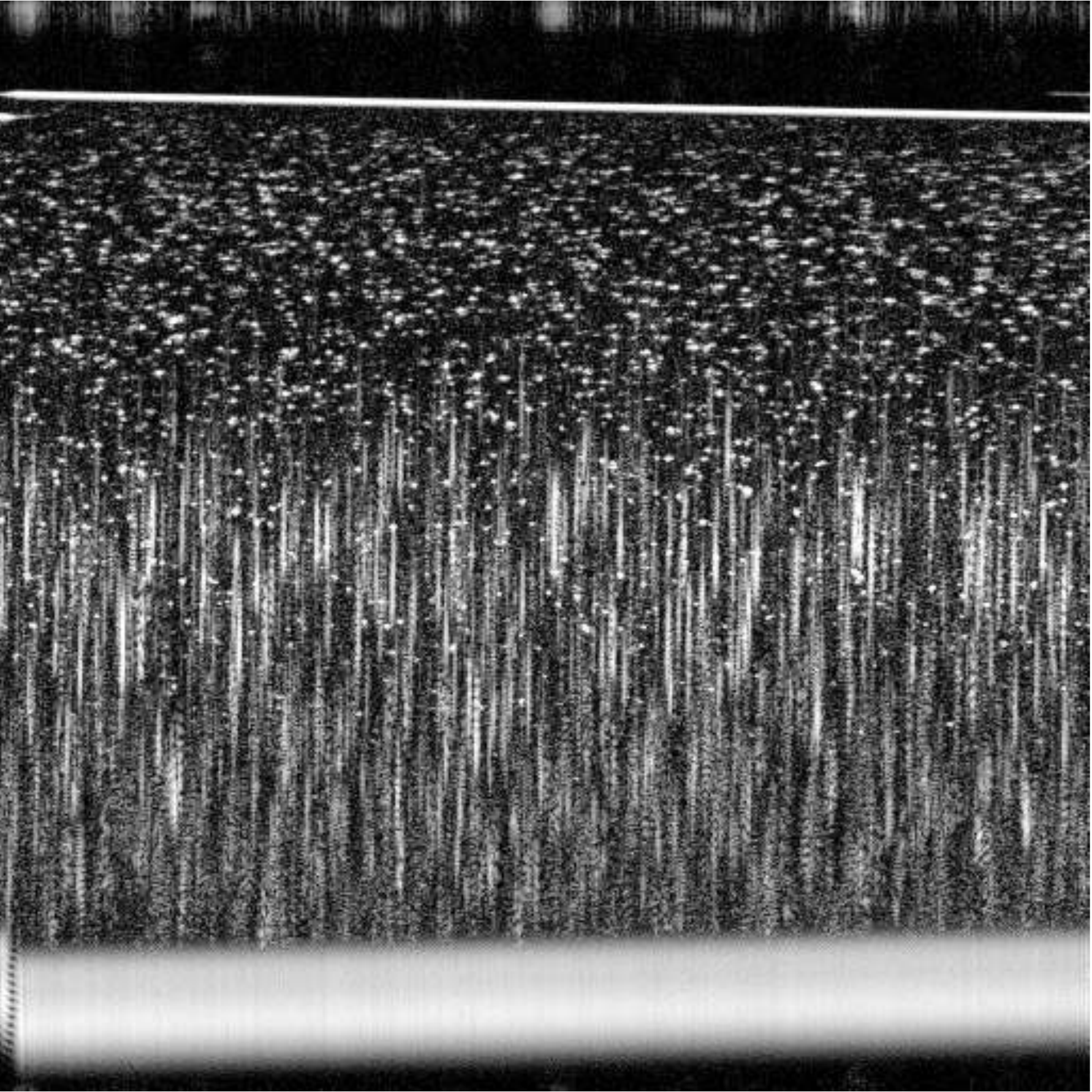}
				\includegraphics[width=\textwidth]{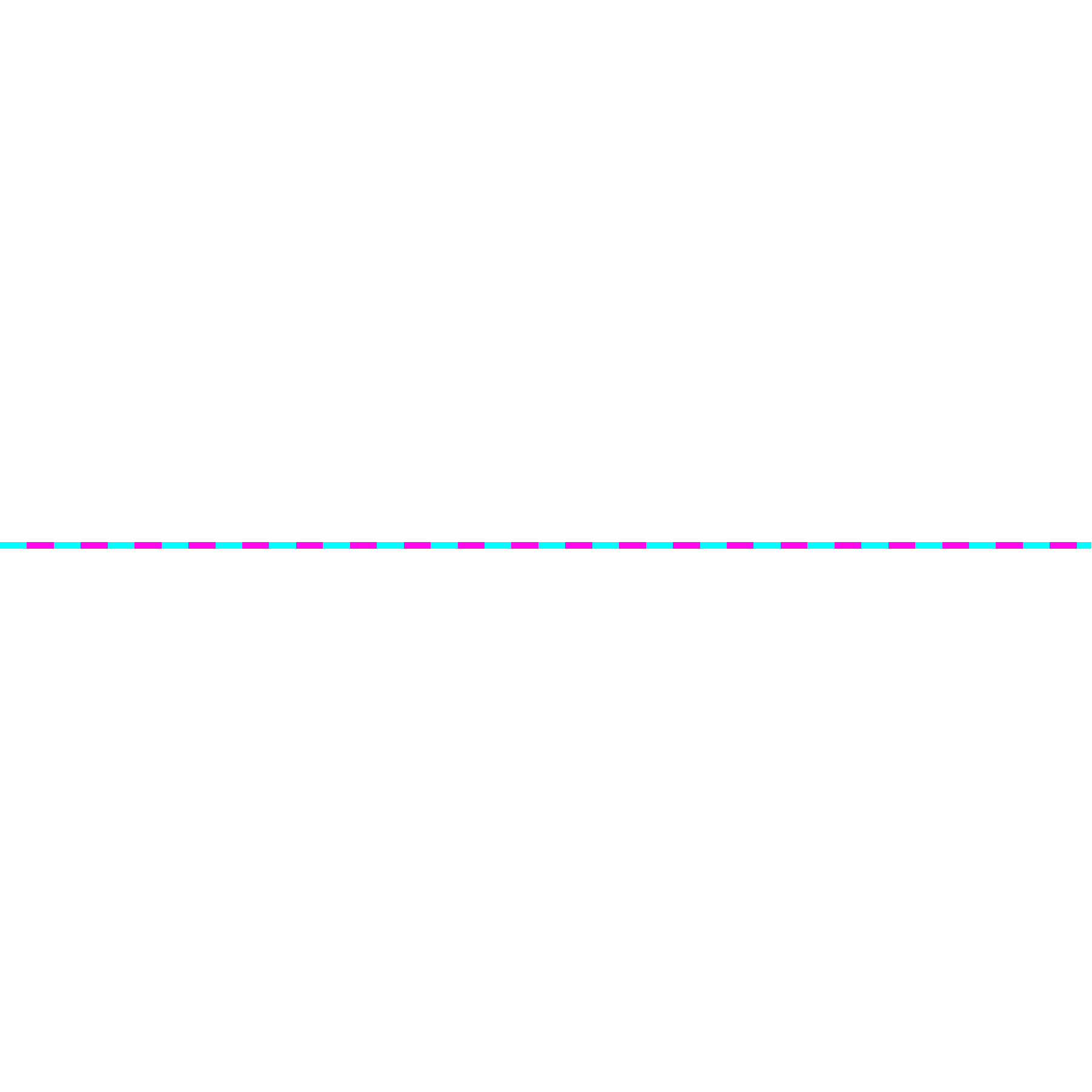} 
			\end{overpic}
			\caption{ISAM}
		\end{subfigure}
		\begin{subfigure}[b]{0.32\textwidth}
			\begin{overpic}[width=\textwidth]{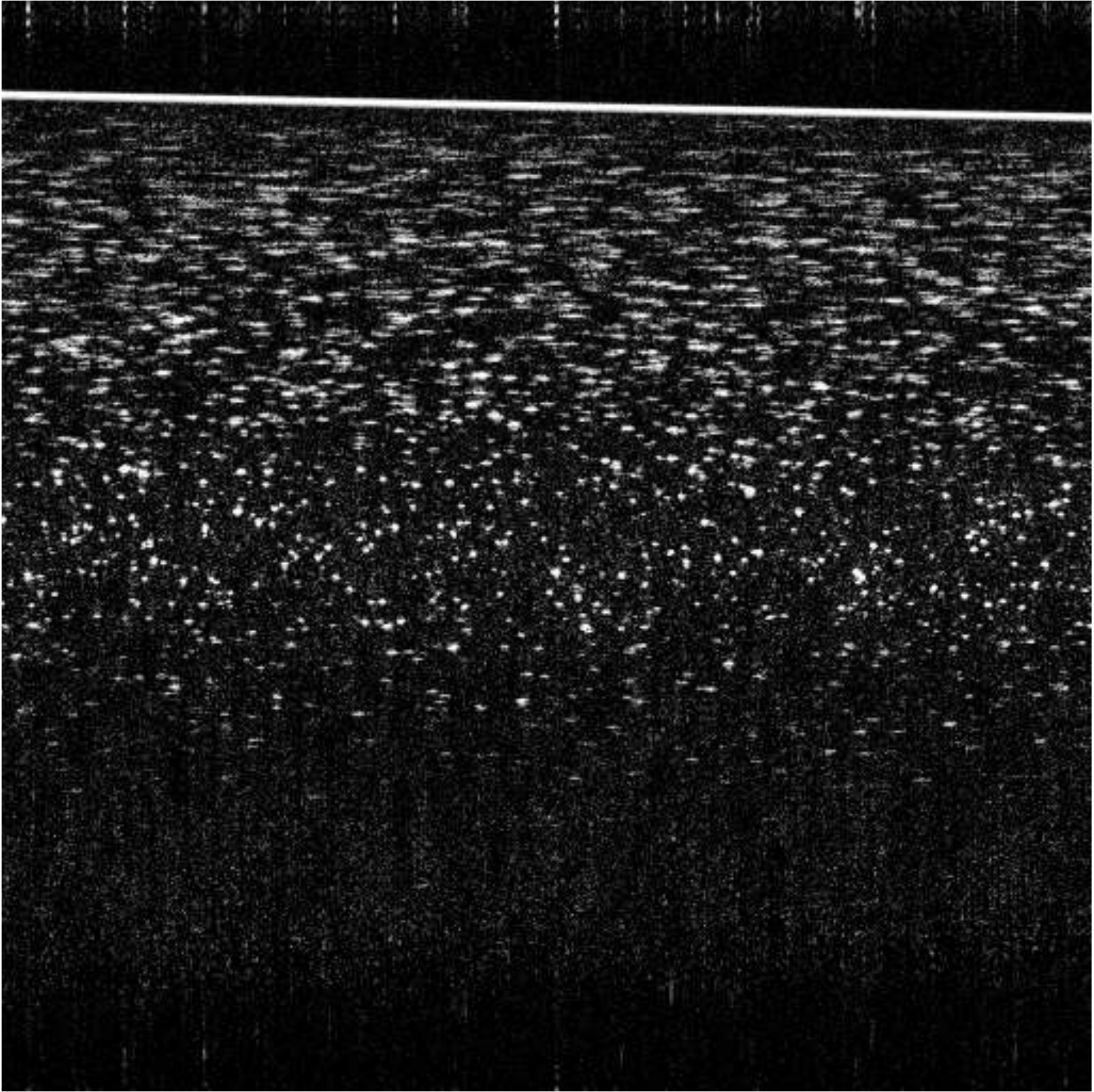}
				\includegraphics[width=\textwidth]{focal_overlay_3} 
			\end{overpic}
			\caption{DEFR}
		\end{subfigure}
		\begin{subfigure}[b]{0.32\textwidth}
			\begin{overpic}[width=\textwidth]{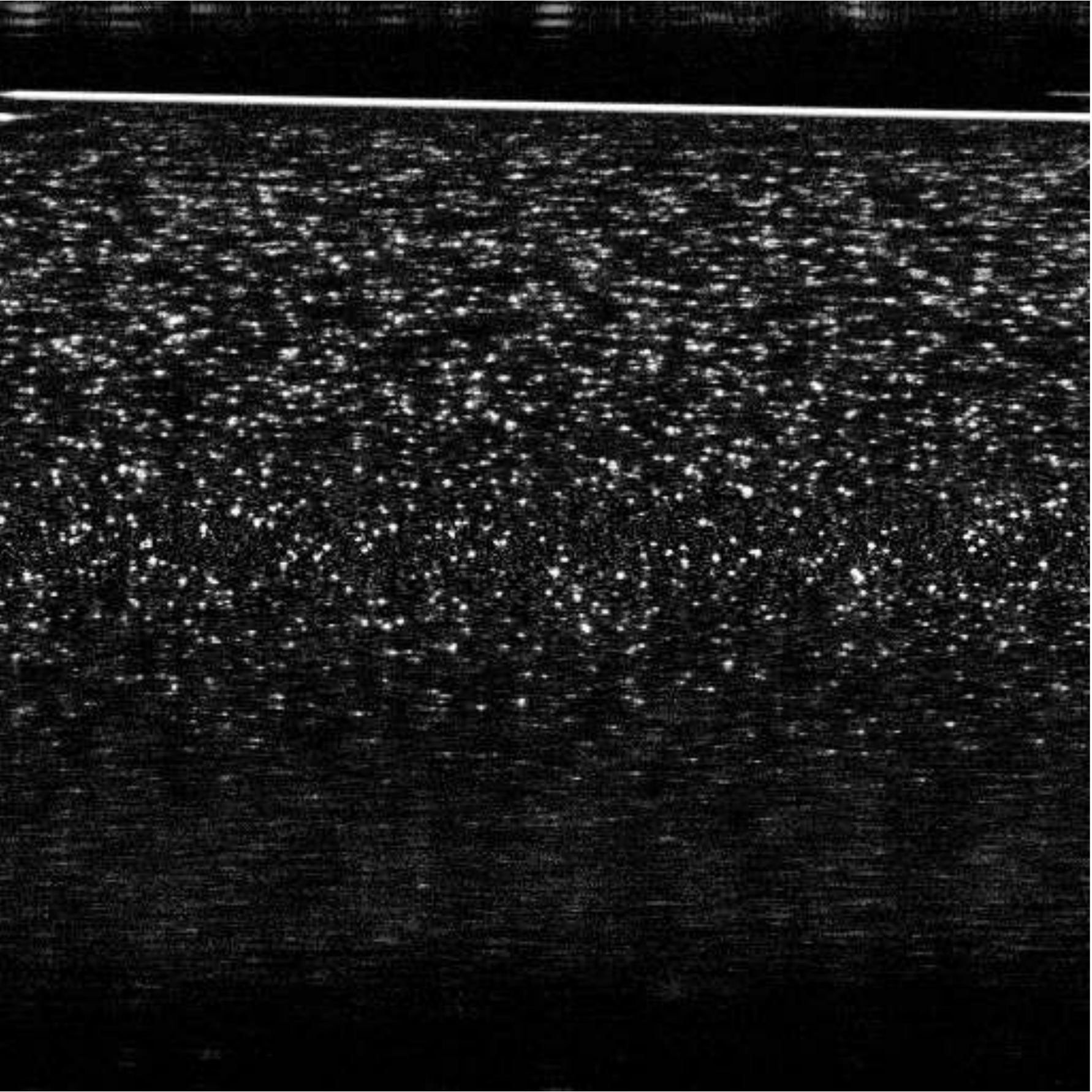}
				\includegraphics[width=\textwidth]{focal_overlay_3} 
			\end{overpic}
			\caption{DEFR+ISAM}
		\end{subfigure}
		\begin{subfigure}[b]{0.32\textwidth}
			\begin{overpic}[width=\textwidth]{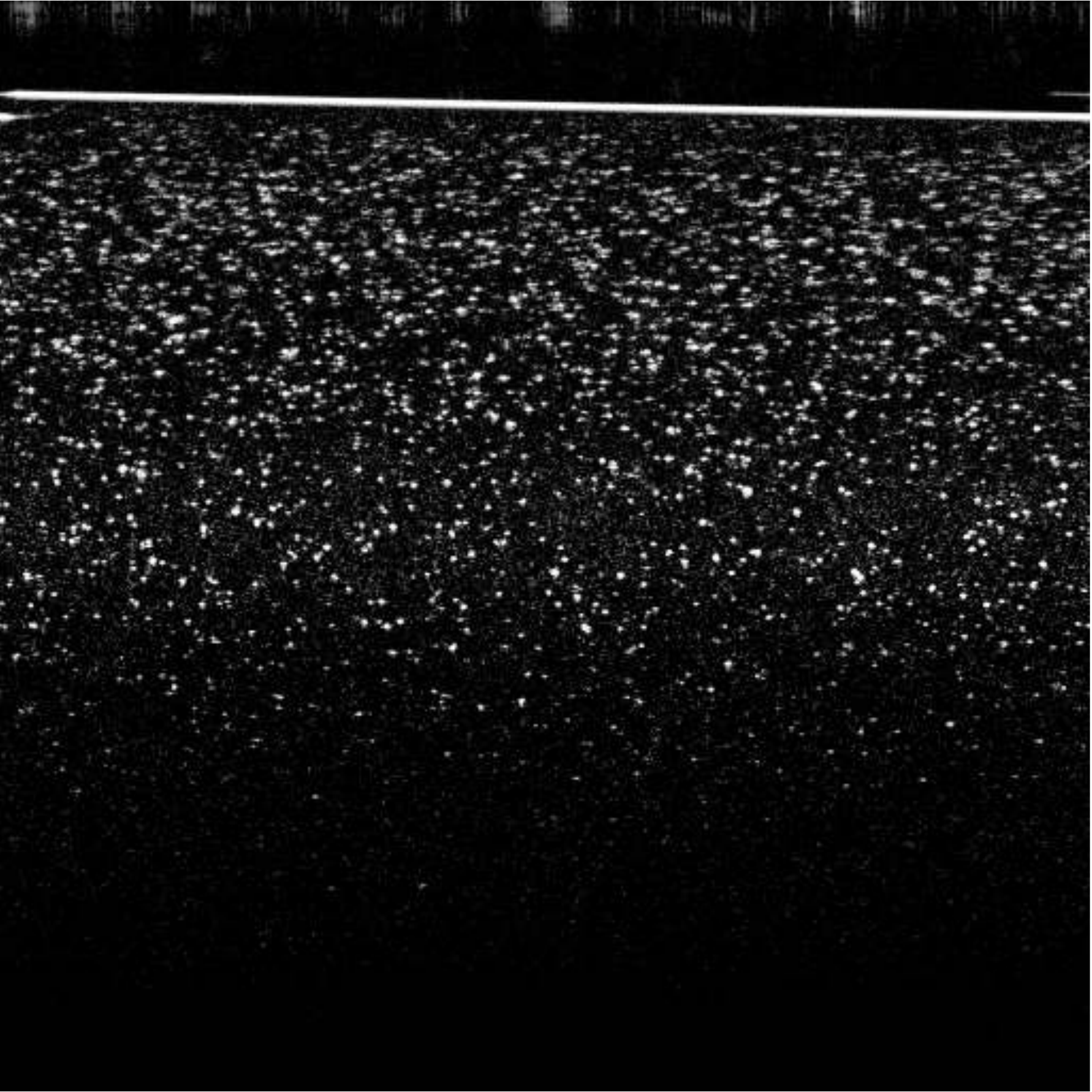}
				\includegraphics[width=\textwidth]{focal_overlay_3} 
			\end{overpic}
			\caption{MBIR+}
		\end{subfigure}
		\caption{\label{fig:quant_ti}Reconstructions from synthetic beaded gel data with ground truth and measurements produced as shown in Fig.~\ref{fig:flow}. The methods and their implementation are detailed in Sec.~\ref{sec:methods}.}
	\end{figure}
	
	\begin{figure}[!htbp]
		\centering
		\begin{subfigure}[b]{0.32\textwidth}
			\begin{overpic}[width=\textwidth]{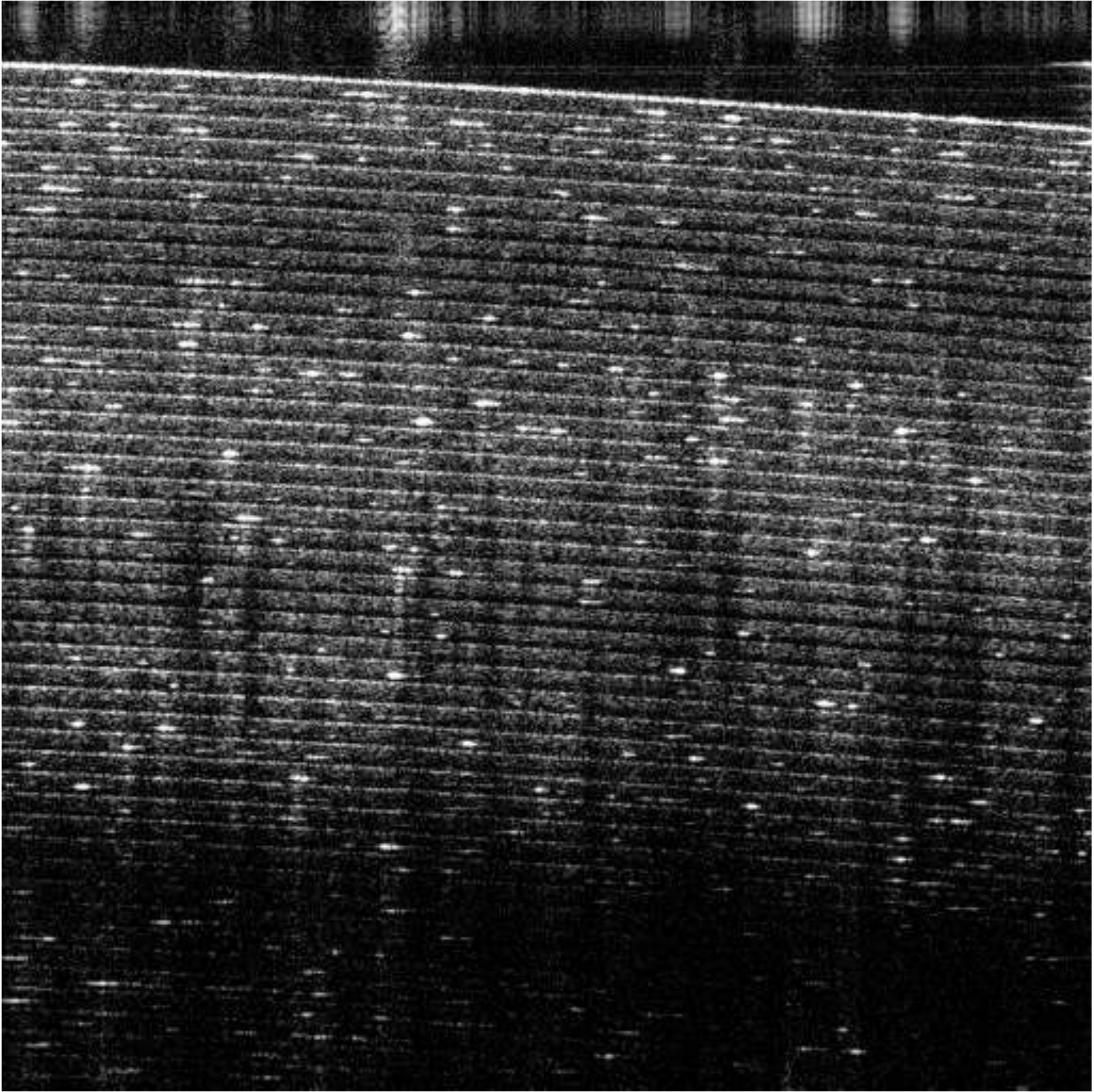}
				\includegraphics[width=\textwidth]{focal_overlay_2} 
			\end{overpic}
			\caption{ground truth}
		\end{subfigure}
		\begin{subfigure}[b]{0.32\textwidth}
			\begin{overpic}[width=\textwidth]{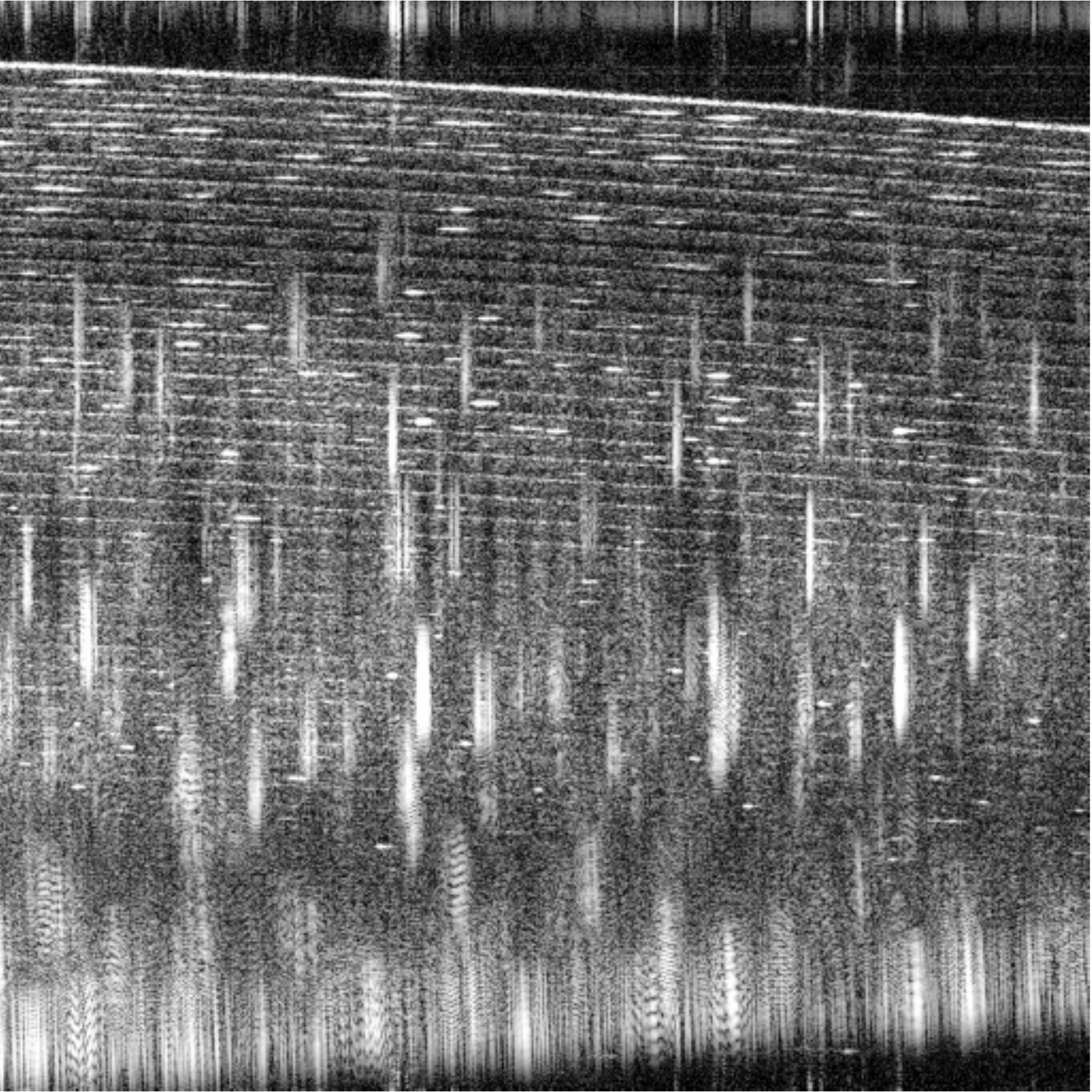}
				\includegraphics[width=\textwidth]{focal_overlay_1} 
			\end{overpic}
			\caption{direct IFFT}
		\end{subfigure}
		\begin{subfigure}[b]{0.32\textwidth}
			\begin{overpic}[width=\textwidth]{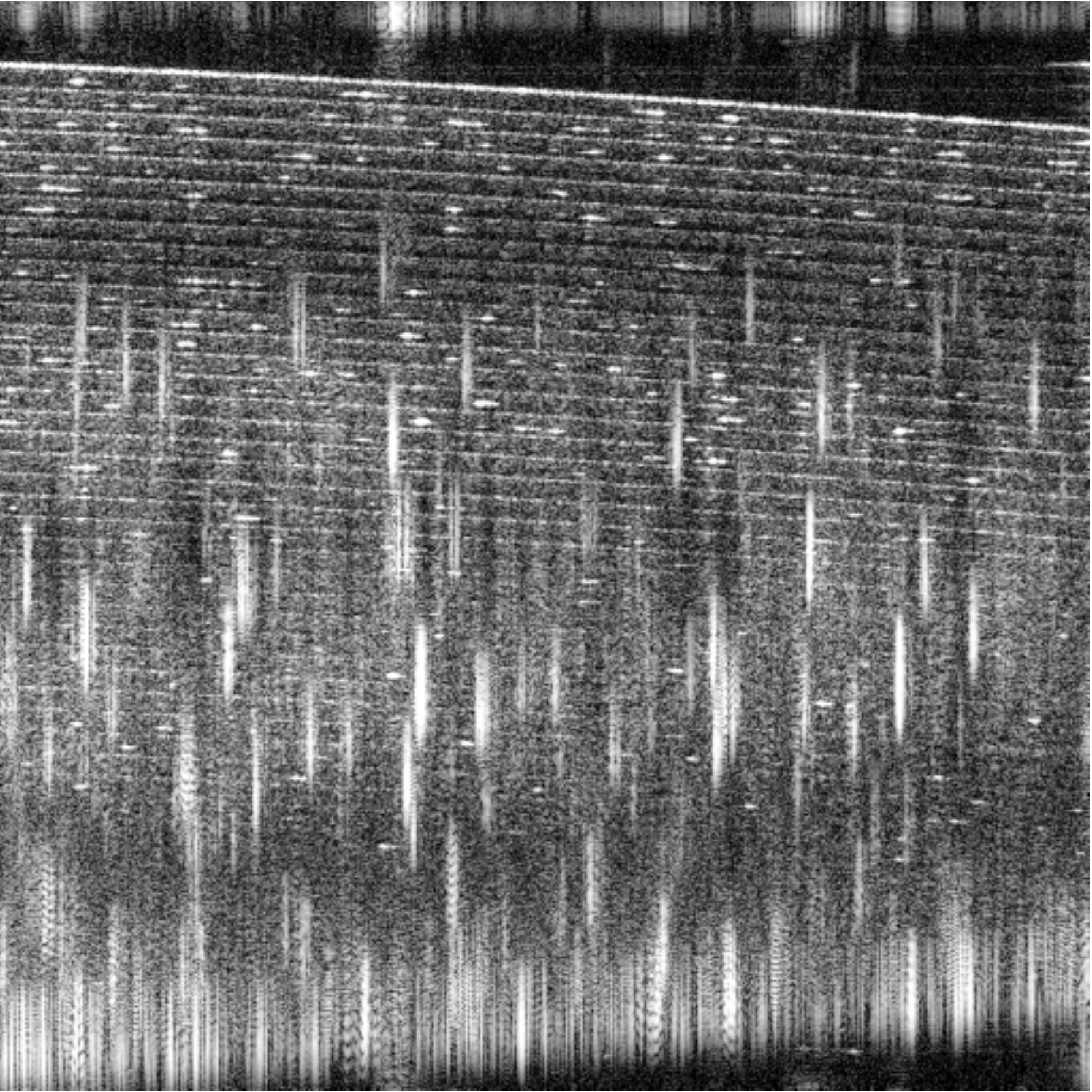}
				\includegraphics[width=\textwidth]{focal_overlay_3} 
			\end{overpic}
			\caption{ISAM}
		\end{subfigure}
		\begin{subfigure}[b]{0.32\textwidth}
			\begin{overpic}[width=\textwidth]{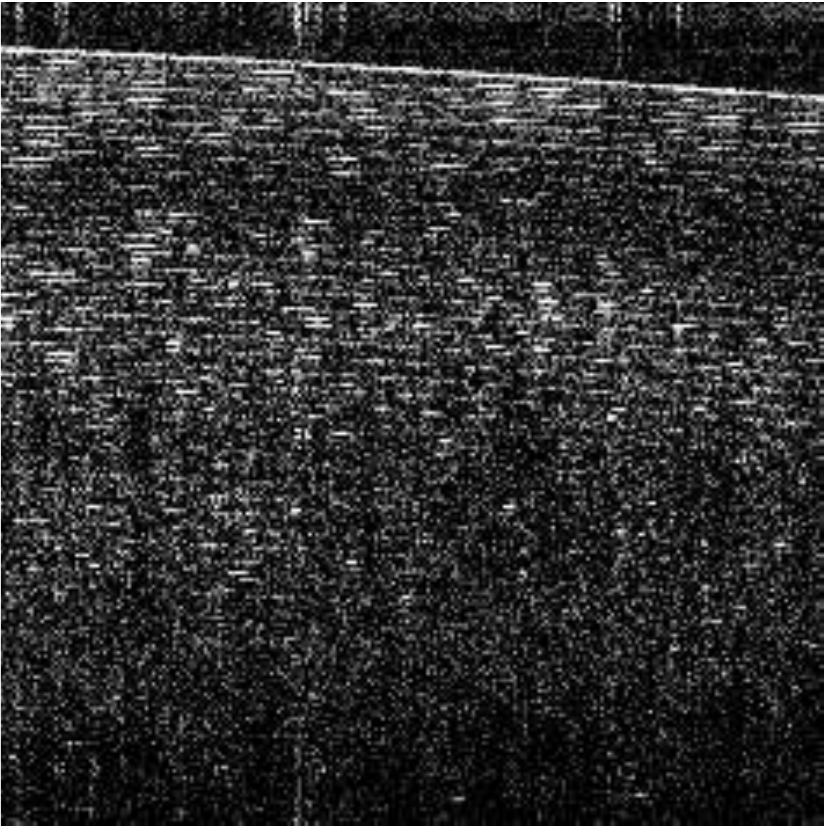}
				\includegraphics[width=\textwidth]{focal_overlay_3} 
			\end{overpic}
			\caption{DEFR}
		\end{subfigure}
		\begin{subfigure}[b]{0.32\textwidth}
			\begin{overpic}[width=\textwidth]{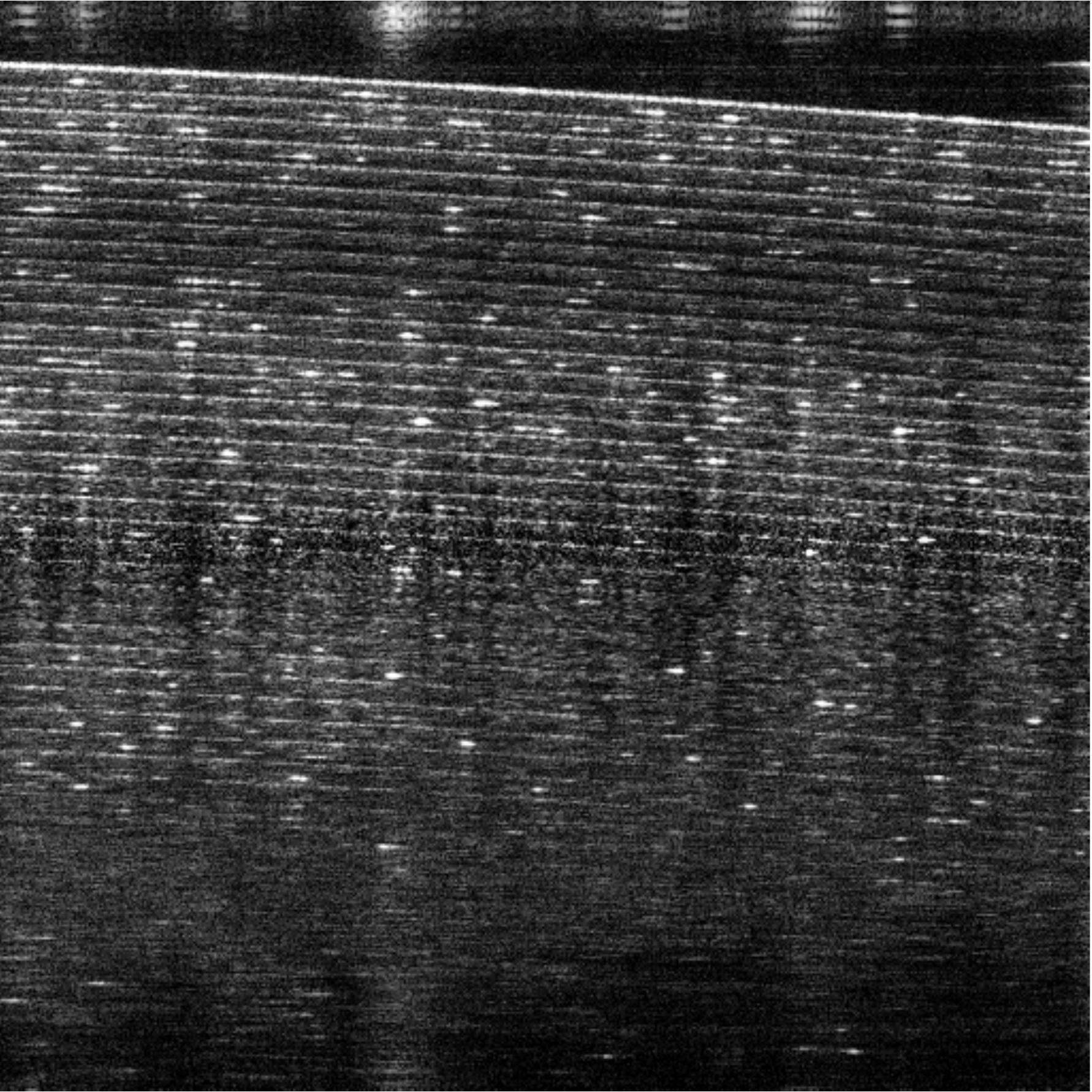}
				\includegraphics[width=\textwidth]{focal_overlay_3} 
			\end{overpic}
			\caption{DEFR+ISAM}
		\end{subfigure}
		\begin{subfigure}[b]{0.32\textwidth}
			\begin{overpic}[width=\textwidth]{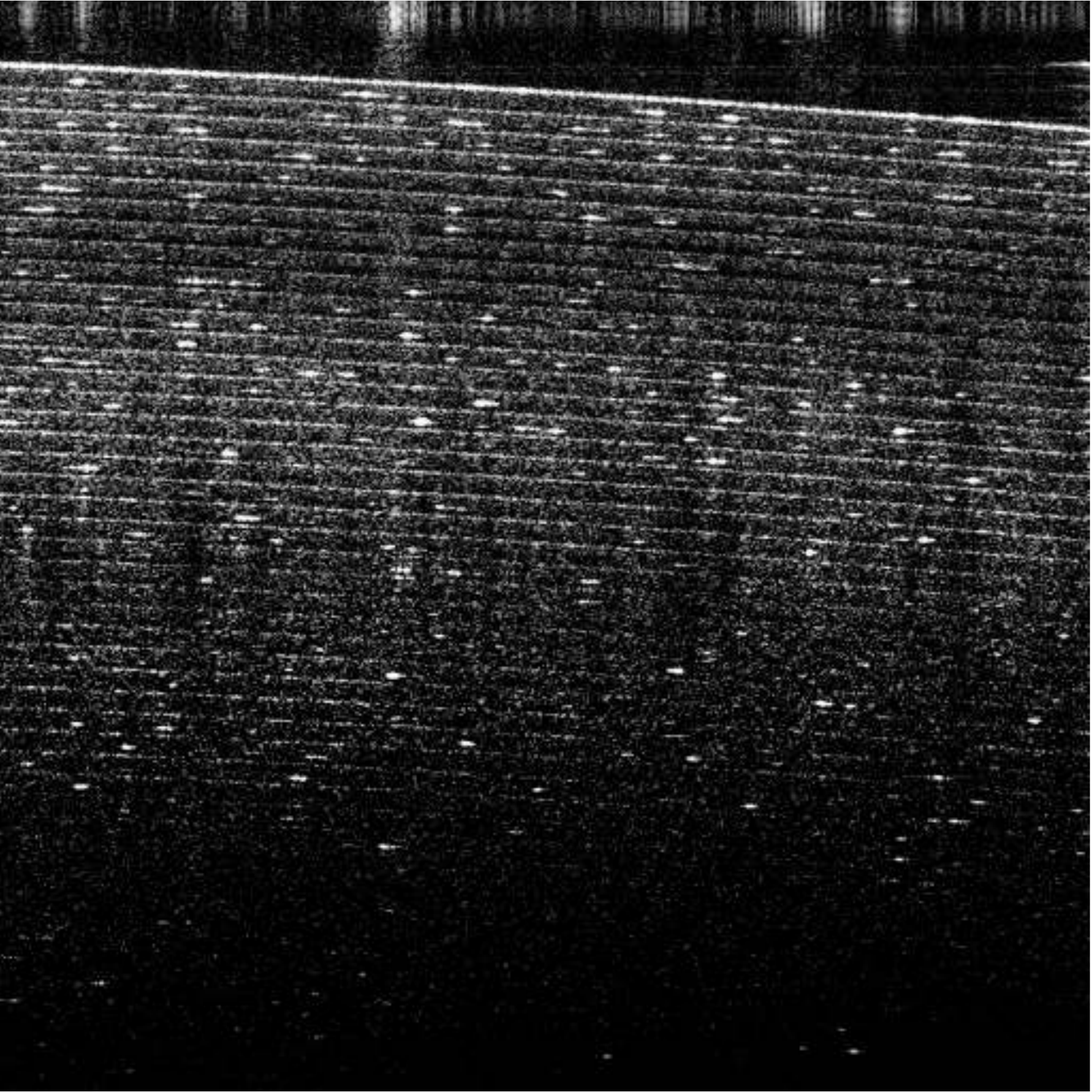}
				\includegraphics[width=\textwidth]{focal_overlay_3} 
			\end{overpic}
			\caption{MBIR+}
		\end{subfigure}
		\caption{\label{fig:quant_tape}Reconstructions from synthetic Scotch tape data with ground truth and measurements produced as shown in Fig.~\ref{fig:flow}. The methods and their implementation are detailed in Sec.~\ref{sec:methods}.}
	\end{figure}
	
	\begin{figure}[!htbp]
		\centering
		\begin{subfigure}[b]{0.32\textwidth}
			\begin{overpic}[width=\textwidth]{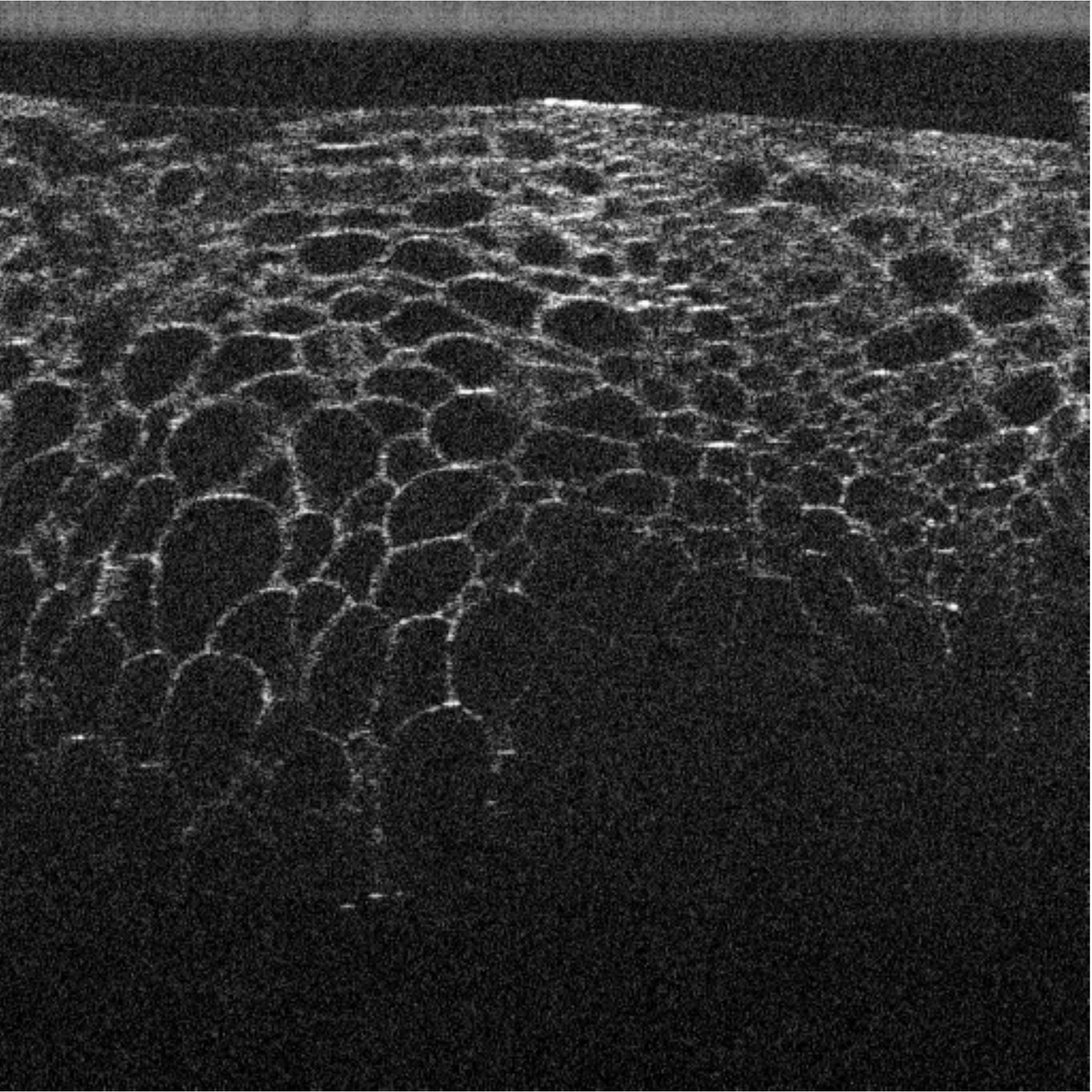}
				\includegraphics[width=\textwidth]{focal_overlay_2} 
			\end{overpic}
			\caption{ground truth}
		\end{subfigure}
		\begin{subfigure}[b]{0.32\textwidth}
			\begin{overpic}[width=\textwidth]{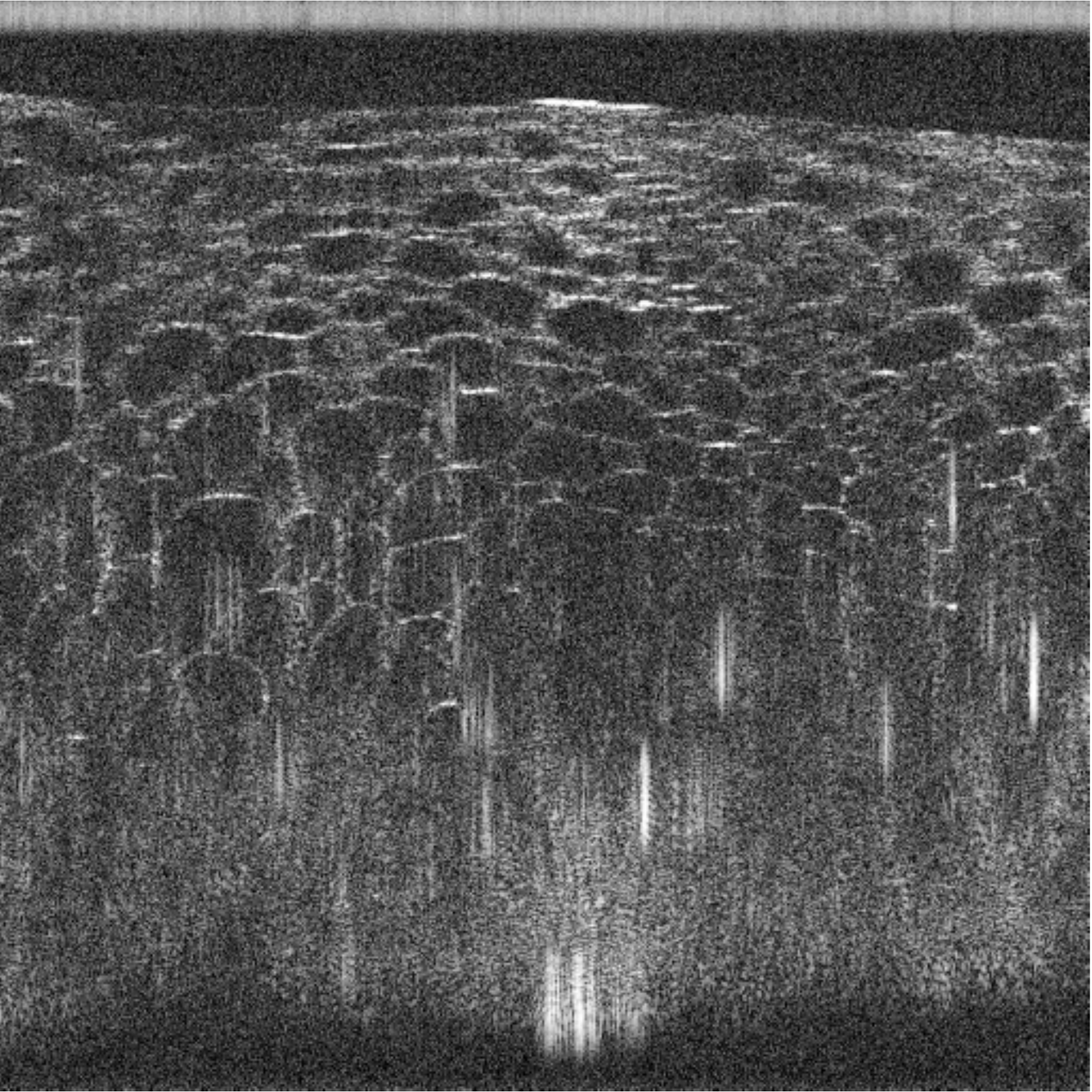}
				\includegraphics[width=\textwidth]{focal_overlay_1} 
			\end{overpic}
			\caption{direct IFFT}
		\end{subfigure}
		\begin{subfigure}[b]{0.32\textwidth}
			\begin{overpic}[width=\textwidth]{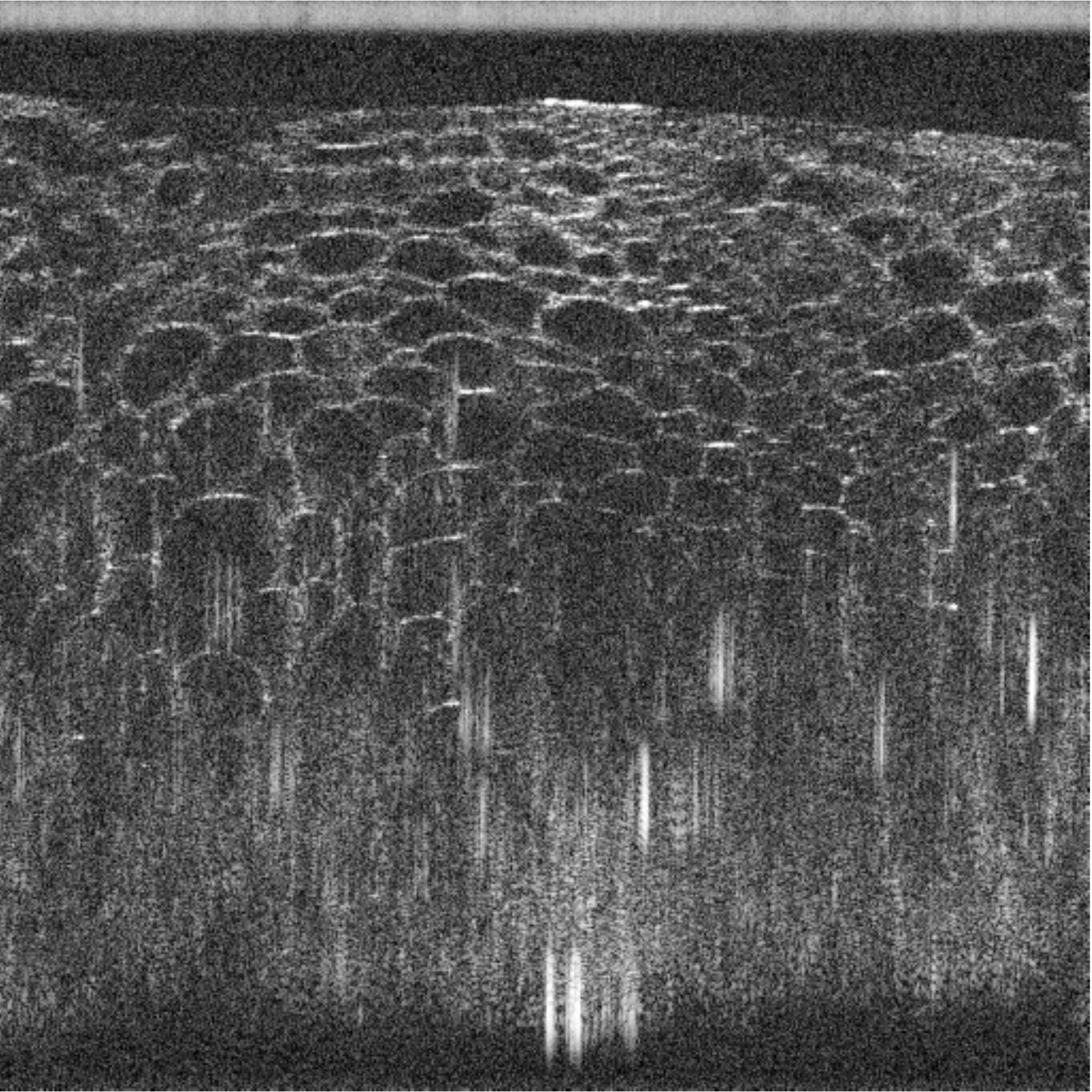}
				\includegraphics[width=\textwidth]{focal_overlay_3} 
			\end{overpic}
			\caption{ISAM}
		\end{subfigure}
		\begin{subfigure}[b]{0.32\textwidth}
			\begin{overpic}[width=\textwidth]{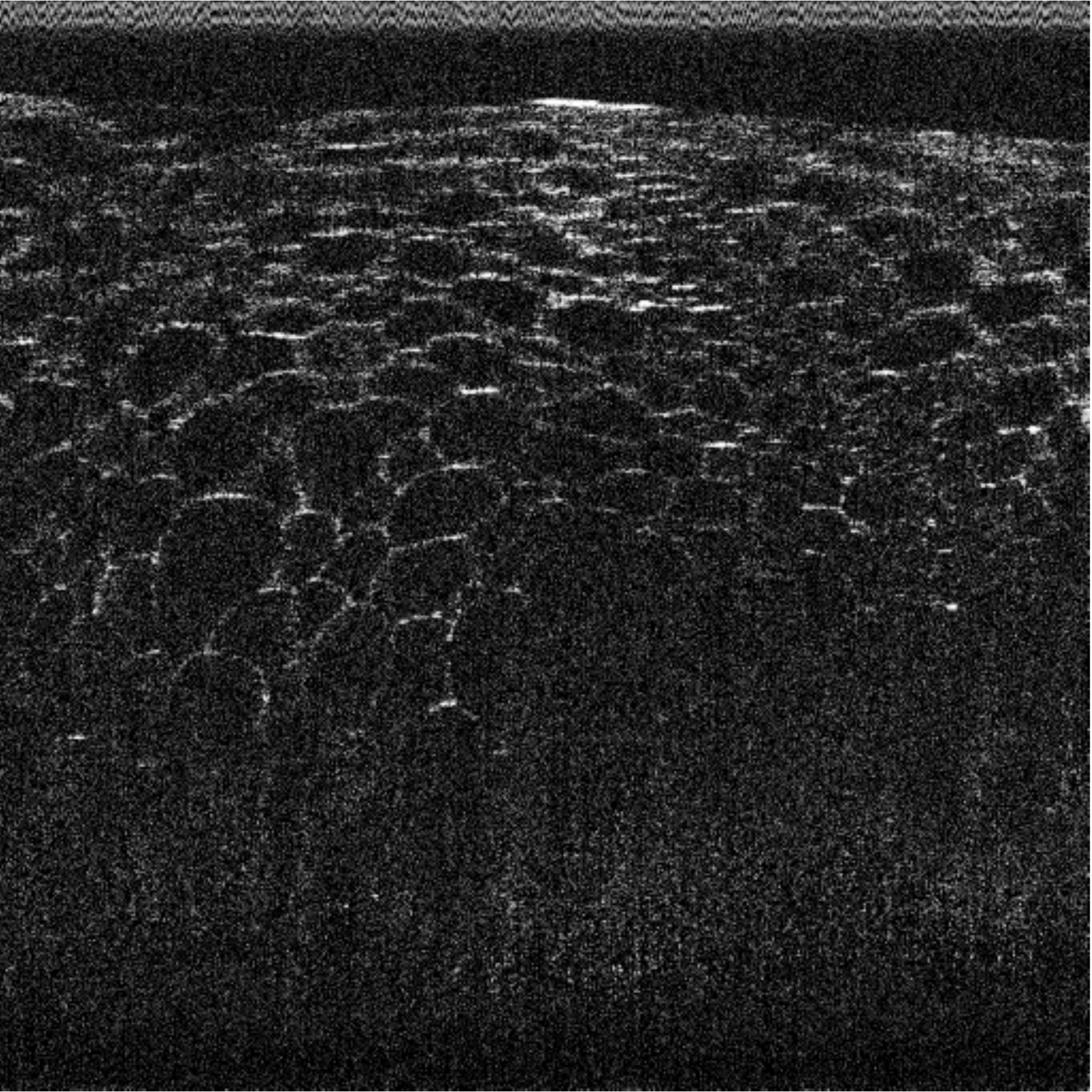}
				\includegraphics[width=\textwidth]{focal_overlay_3} 
			\end{overpic}
			\caption{DEFR}
		\end{subfigure}
		\begin{subfigure}[b]{0.32\textwidth}
			\begin{overpic}[width=\textwidth]{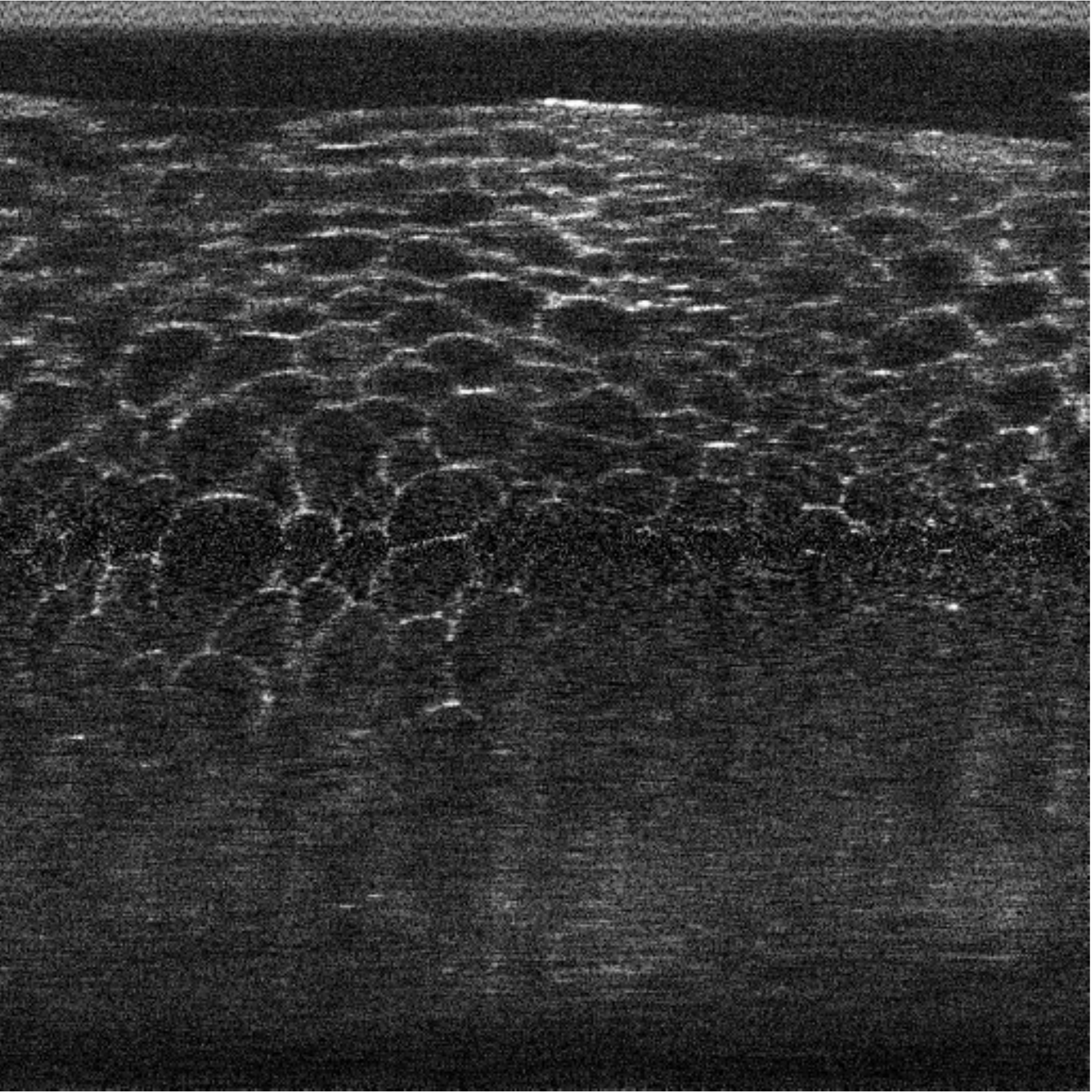}
				\includegraphics[width=\textwidth]{focal_overlay_3} 
			\end{overpic}
			\caption{DEFR+ISAM}
		\end{subfigure}
		\begin{subfigure}[b]{0.32\textwidth}
			\begin{overpic}[width=\textwidth]{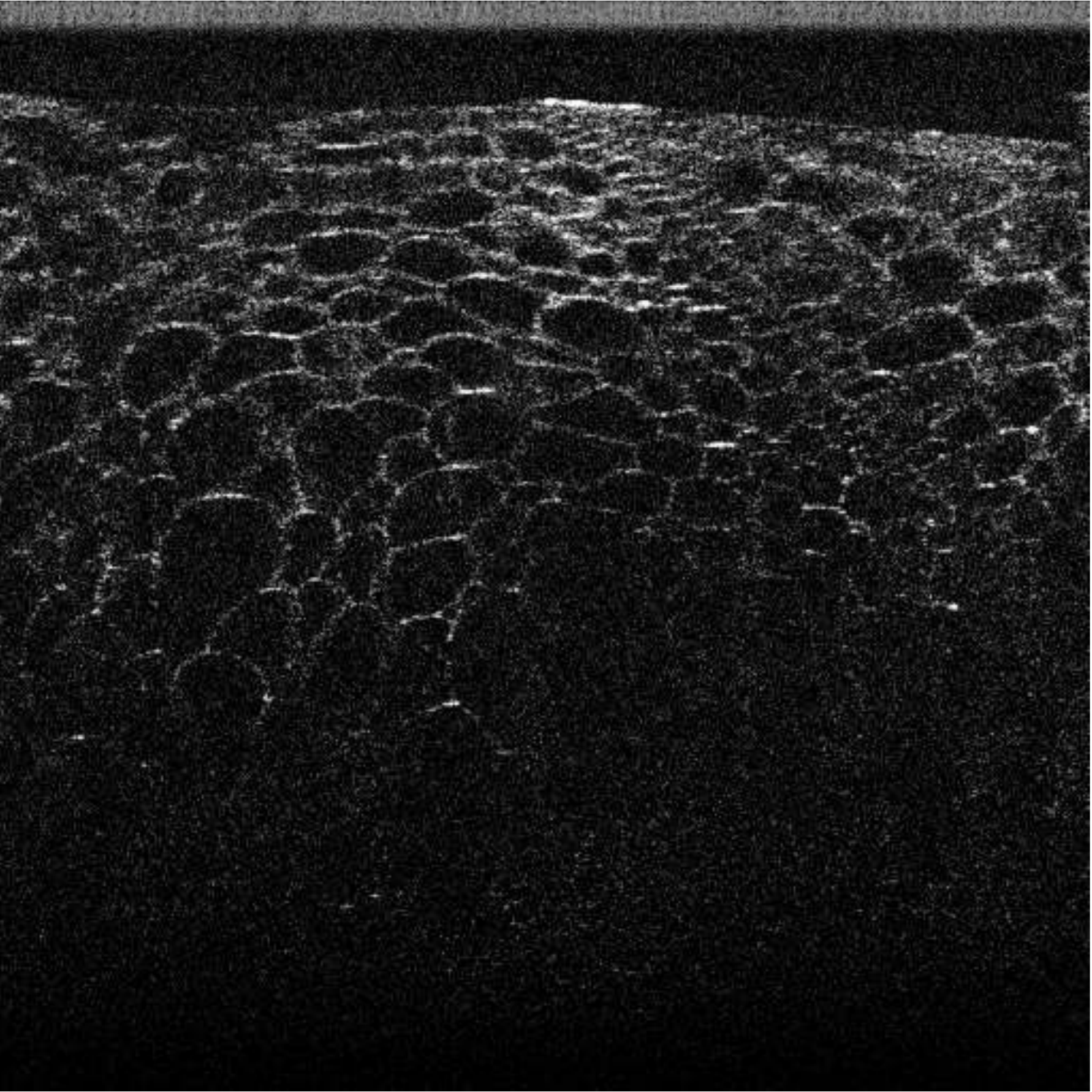}
				\includegraphics[width=\textwidth]{focal_overlay_3} 
			\end{overpic}
			\caption{MBIR+}
		\end{subfigure}
		\caption{\label{fig:quant_cucumber}Reconstructions from synthetic cucumber data with ground truth and measurements produced as shown in Fig.~\ref{fig:flow}. The methods and their implementation are detailed in Sec.~\ref{sec:methods}.}
	\end{figure}
	
	\begin{table}[!htbp]
		\centering
		\caption{\label{tab:quant}\textbf{Quantitative results from synthetic full-range data (best method in bold)}.}
		\begin{tabular}{c|c|c|c|c|c|c|c}
			sample & metric & direct & ISAM & DEFR & DEFR+ISAM & MBIR & MBIR+ \\
			\hline
			\multirow{ 3}{*}{beaded gel}&RMSE&5.23&3.02&4.27&0.797&0.648&\textbf{0.548}\\
			&PSNR&8.16&8.65&15.5&18.7&19.1&\textbf{21.3}\\
			&SSIM&0.147&0.341&0.199&0.371&0.370&\textbf{0.595}\\
			\hline
			\multirow{ 3}{*}{Scotch tape}&RMSE&4.61&2.68&3.76&1.05&0.874&\textbf{0.777}\\
			&PSNR&9.11&9.70&12.8&15.1&14.8&\textbf{16.0}\\
			&SSIM&0.178&0.365&0.168&0.314&0.349&\textbf{0.411}\\
			\hline
			\multirow{ 3}{*}{cucumber}&RMSE&0.860&0.493&0.704&0.388&0.295&\textbf{0.258}\\
			&PSNR&16.7&18.0&19.2&22.4&23.1&\textbf{25.1}\\
			&SSIM&0.105&0.315&0.147&0.332&0.369&\textbf{0.517}\\
			\hline
		\end{tabular}
	\end{table}
	
	The samples used in the quantitative study were the beaded gel, tape and cucumber, as described in Sec.~\ref{sec:mat}, although with the focal plane positioned at half the positive delay range. Reconstructions are shown in Fig.~\ref{fig:quant_ti}, Fig.~\ref{fig:quant_tape} and Fig.~\ref{fig:quant_tape}, with results of the quantitative analysis presented in Tab.~\ref{tab:quant}. We quantify the root mean squared error (RMSE) between the ground truth and raw reconstructed data. On top of this, we use the peak signal--to--noise ratio (PSNR) and structural similarity index (SSIM) \cite{Wang2004} calculated on the 16-bit images displayed in Fig.~\ref{fig:quant_ti} and Fig.~\ref{fig:quant_tape} after taking the logarithm and scaling the data. We use these additional metrics as they correlate better with human perception of image quality \cite{Wang2004}, and are in the format in which the data is likely to be interpreted.
	
	From the quantitative results in Tab.~\ref{tab:quant}, the performance of our proposed approaches are compelling. Firstly, the simple two-step DEFR+ISAM is effective, and significantly outperforms each of its constituent algorithms, with a significant reduction in error over DEFR for the three samples, but also tends to introduce lateral artifacts in areas from which conjugate components are removed. On top of this, we achieve a 17--24\% reduction in RMSE between DEFR+ISAM and our simple MBIR implementation and a 26--340\% reduction against our MBIR+, which adequately justifies our new algorithm. Although the PSNR and SSIM values between MBIR and DEFR+ISAM are reasonably similar, again our MBIR+ shows a significant gain over both of these, which highlights the benefit of the weighting term.
	
	By comparing the reconstructed images in Fig.~\ref{fig:quant_ti} and Fig.~\ref{fig:quant_tape}, the performance and action of the methods can be seen: DEFR effectively mitigates the conjugate artifacts that are present in the IFFT and ISAM; ISAM brings the areas into focus away from the focal plane that are blurred in IFFT and DEFR; and DEFR+ISAM achieves the combined benefit of its two constituent steps. As well as also having this combined effect, MBIR+ produces images very close to the ground truth, and significantly outperforms any existing model or combination thereof.
	
	The advantages from MBIR approaches over DEFR+ISAM suggests that being able to exploit the multidimensional sparsity present in the focussed image after ISAM resampling is valuable.
	
	\subsection{Real data validation} \label{sec:analysis}
	Whilst the results in Section~\ref{sec:quant} allow us to objectively analyse reconstructions against a ground truth, and between various methods, we also validated this with real full-range measurements. In this section, we apply the same methods to extend the depth range from measurements directly from a commercial OCT system. Since we no longer have a ground truth, this analysis will necessarily be qualitative, with our observations qualified by the results in Sec.~\ref{sec:quant}.
	
	\begin{figure}[!htbp]
		\centering
		\begin{subfigure}[b]{0.32\textwidth}
			\includegraphics[width=\textwidth]{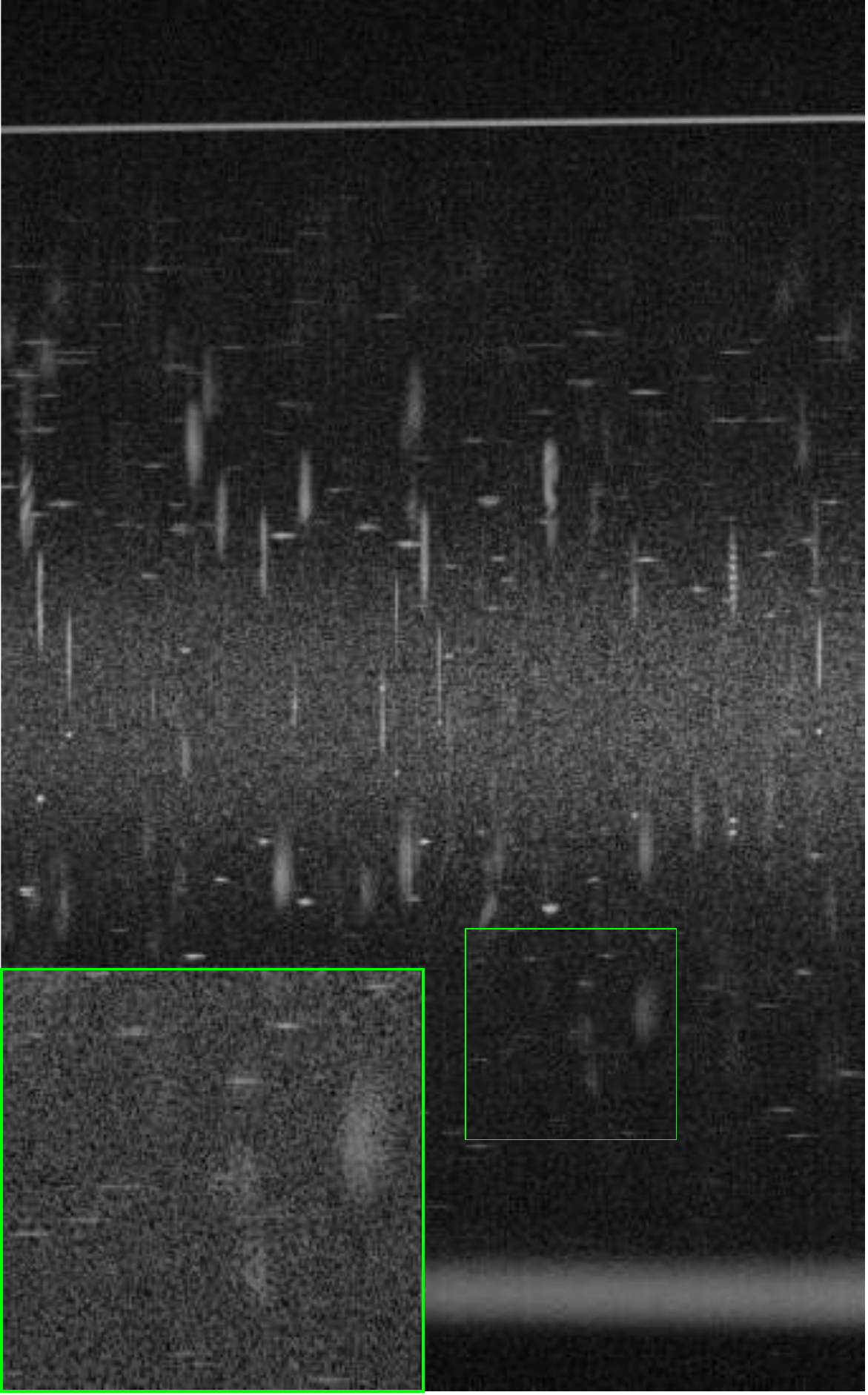}
			\caption{beaded gel direct}
		\end{subfigure}
		\hfil
		\begin{subfigure}[b]{0.32\textwidth}
			\includegraphics[width=\textwidth]{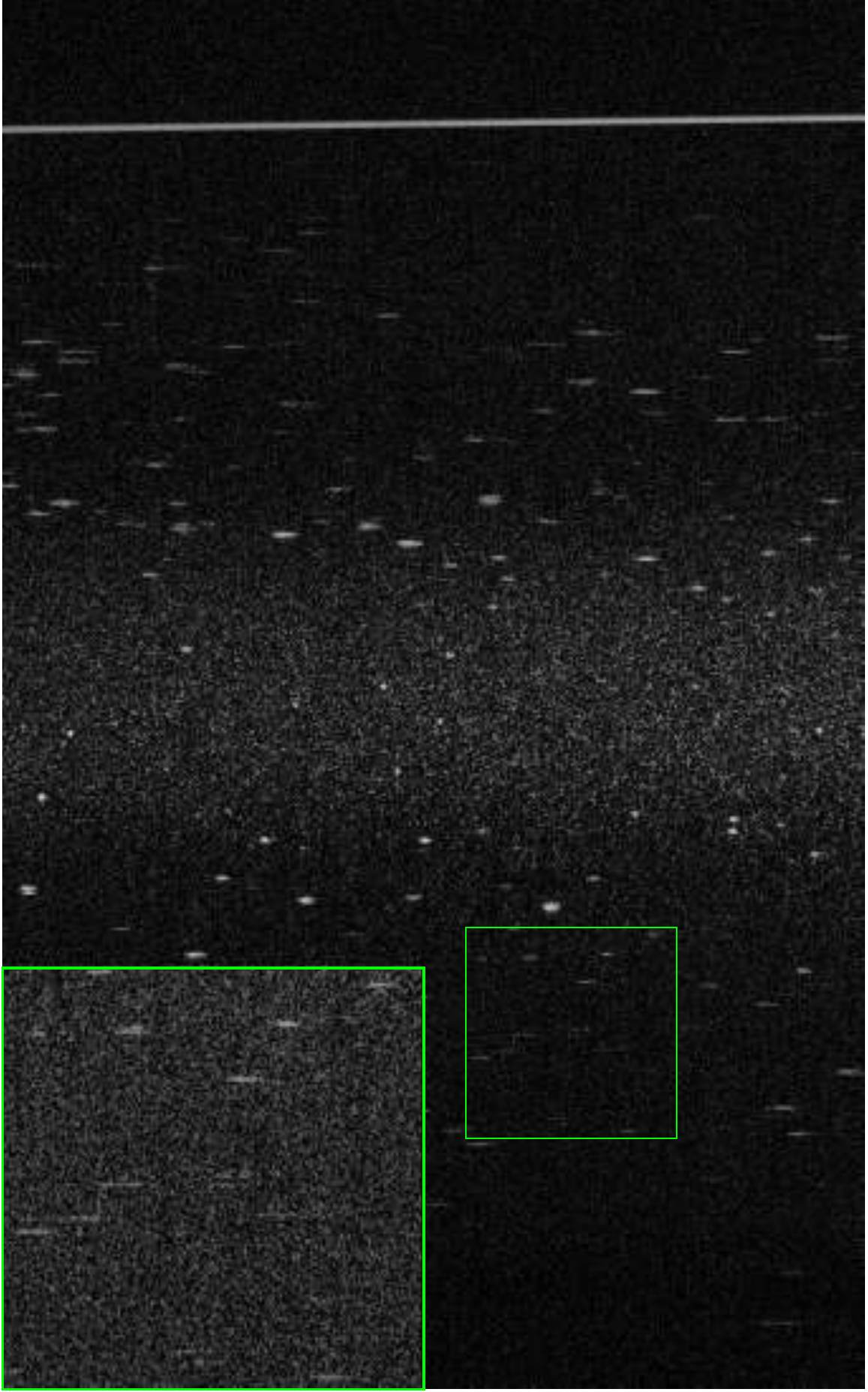}
			\caption{beaded gel DEFR}
		\end{subfigure}
		\hfil
		\begin{subfigure}[b]{0.32\textwidth}
			\includegraphics[width=\textwidth]{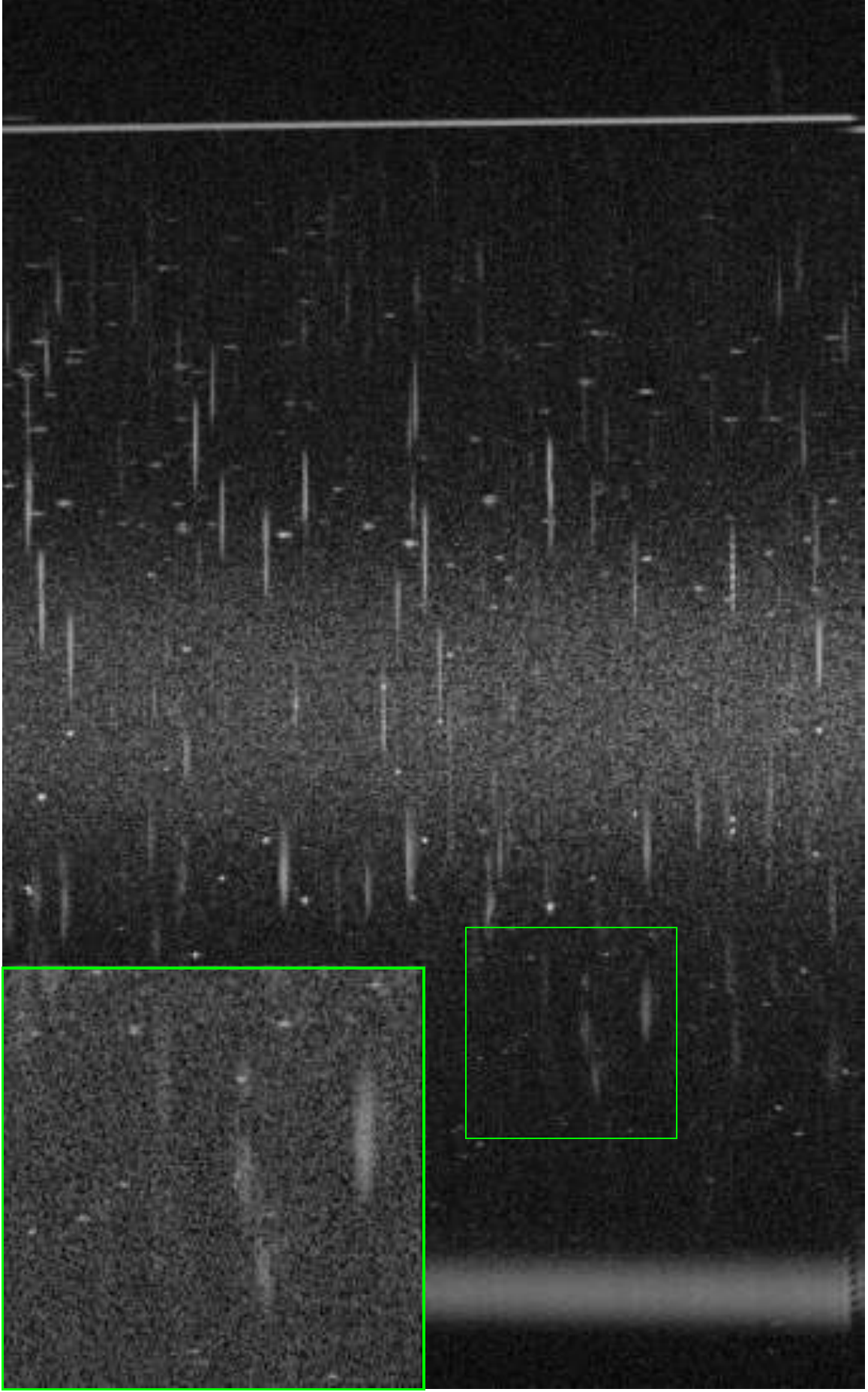}
			\caption{beaded gel ISAM}
		\end{subfigure}
		
		\begin{subfigure}[b]{0.32\textwidth}
			\includegraphics[width=\textwidth]{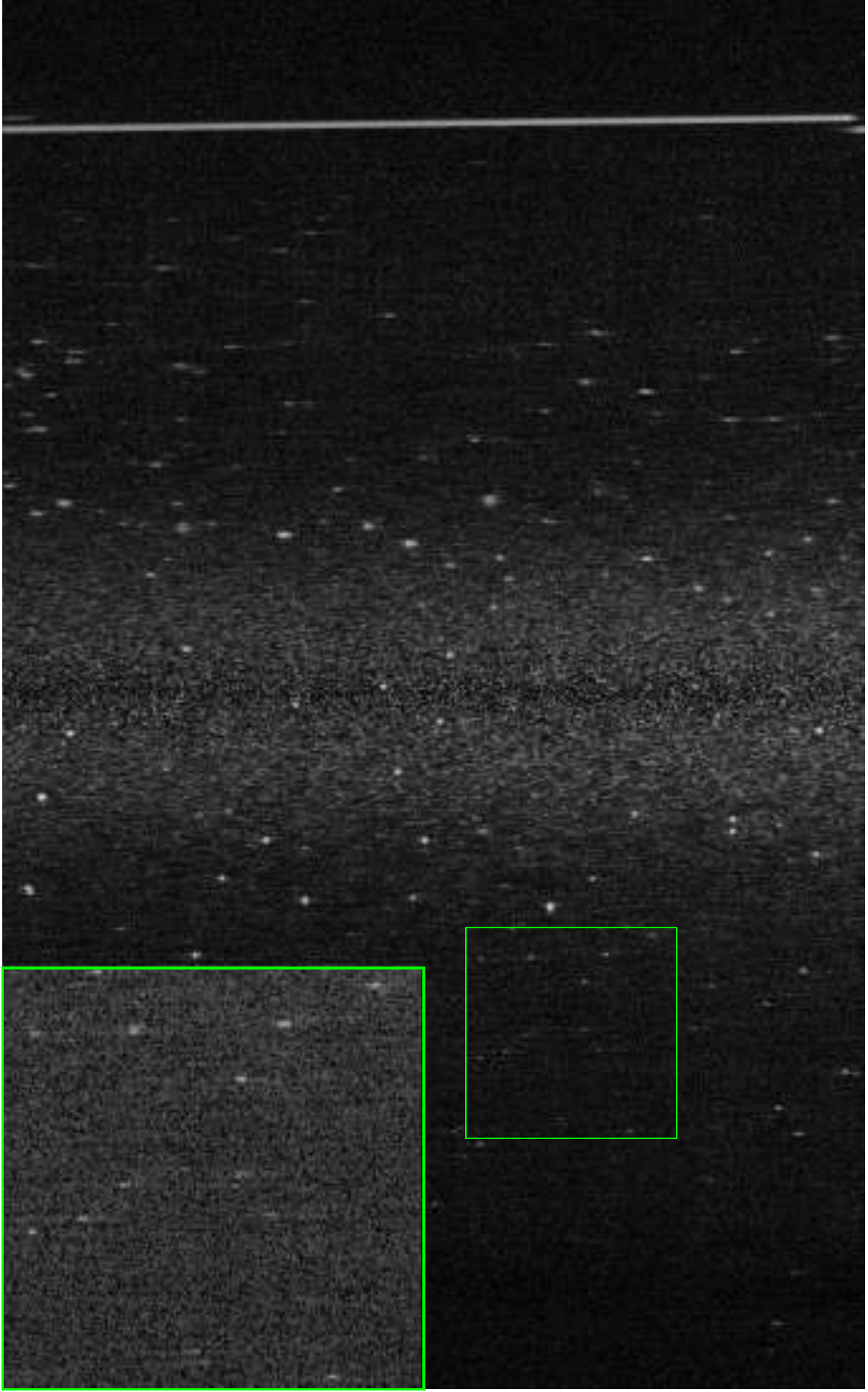}
			\caption{beaded gel DEFR+ISAM}
		\end{subfigure}
		\hfil
		\begin{subfigure}[b]{0.32\textwidth}
			\includegraphics[width=\textwidth]{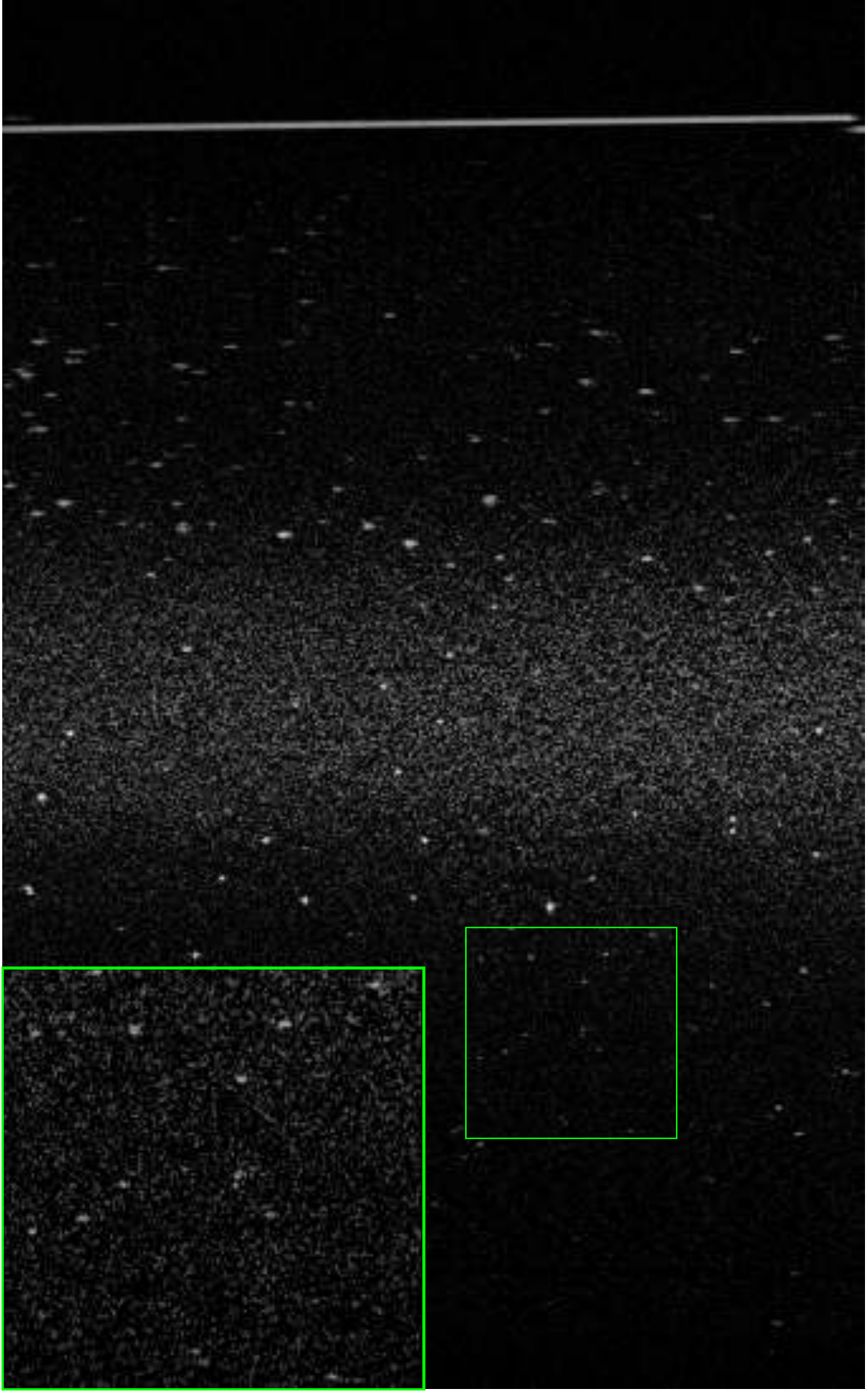}
			\caption{beaded gel MBIR}
		\end{subfigure}
		\hfil
		\begin{subfigure}[b]{0.32\textwidth}
			\includegraphics[width=\textwidth]{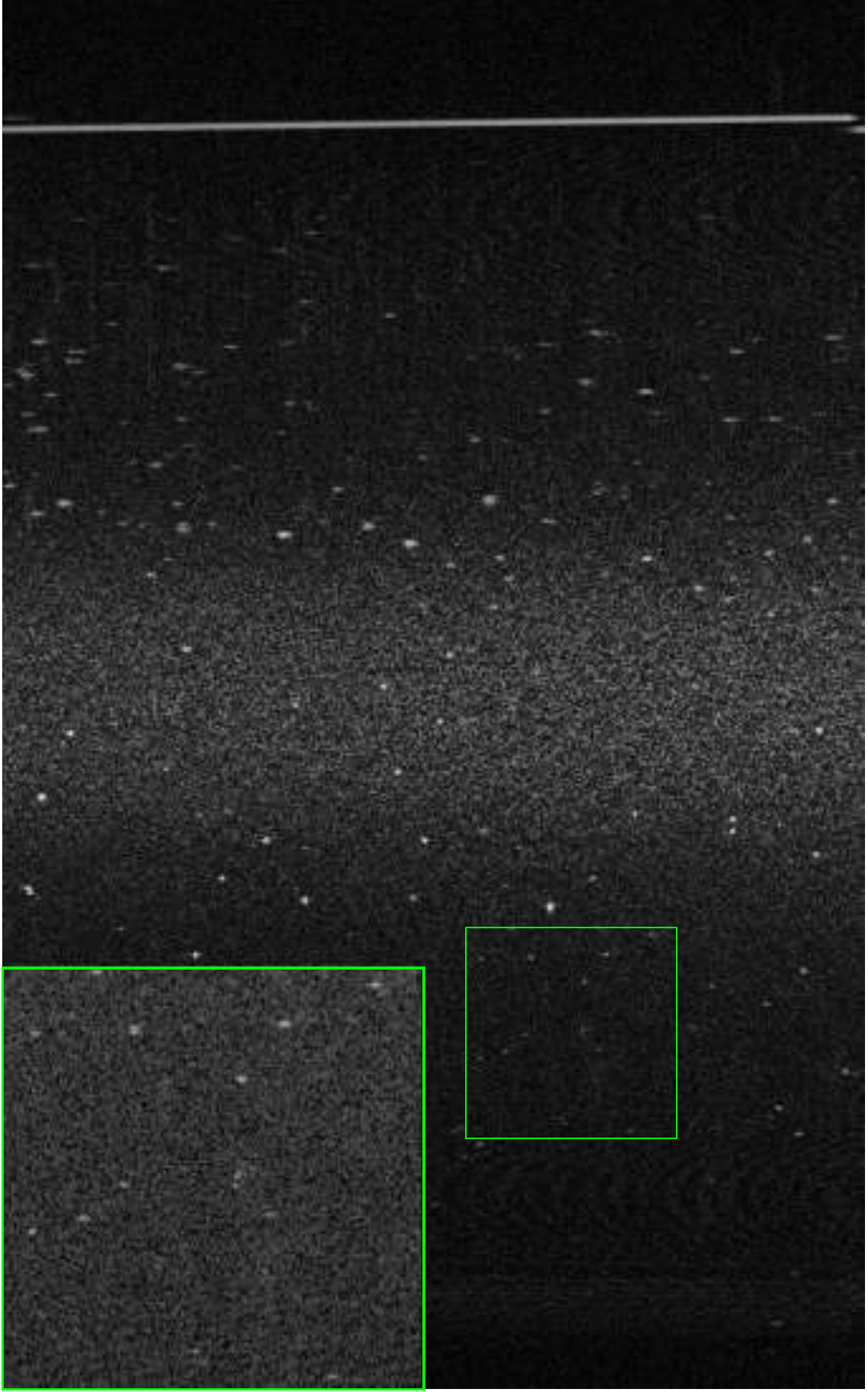}
			\caption{beaded gel MBIR+}
		\end{subfigure}
		\caption{\label{fig:samples_ti}Reconstructions of beaded gel sample with real measurements. The various methods and their implementation are detailed in Sec.~\ref{sec:methods}.}
	\end{figure}
	
	\begin{figure}[!htbp]
		\centering
		\begin{subfigure}[b]{0.32\textwidth}
			\includegraphics[width=\textwidth]{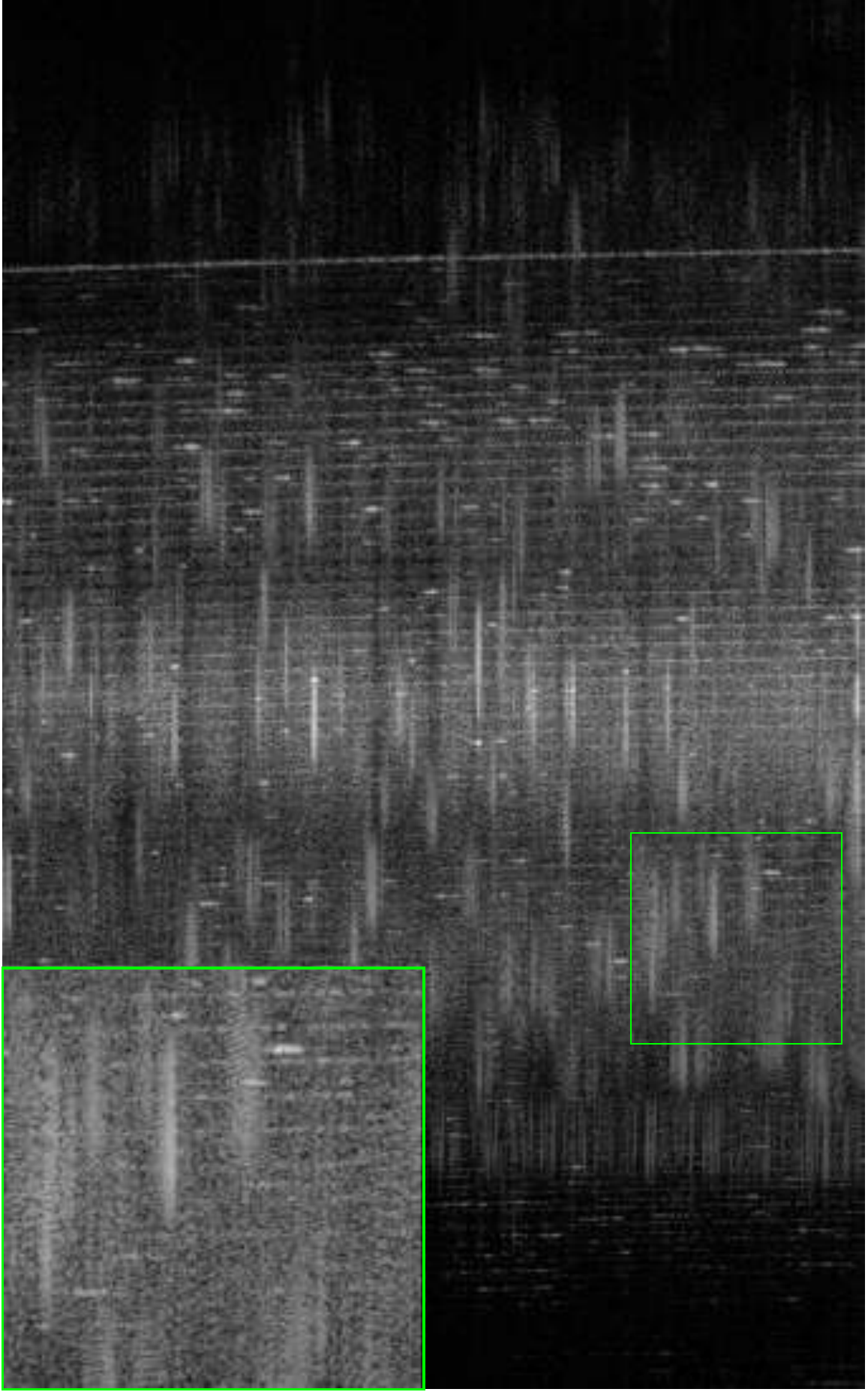}
			\caption{tape direct}
		\end{subfigure}
		\hfil
		\begin{subfigure}[b]{0.32\textwidth}
			\includegraphics[width=\textwidth]{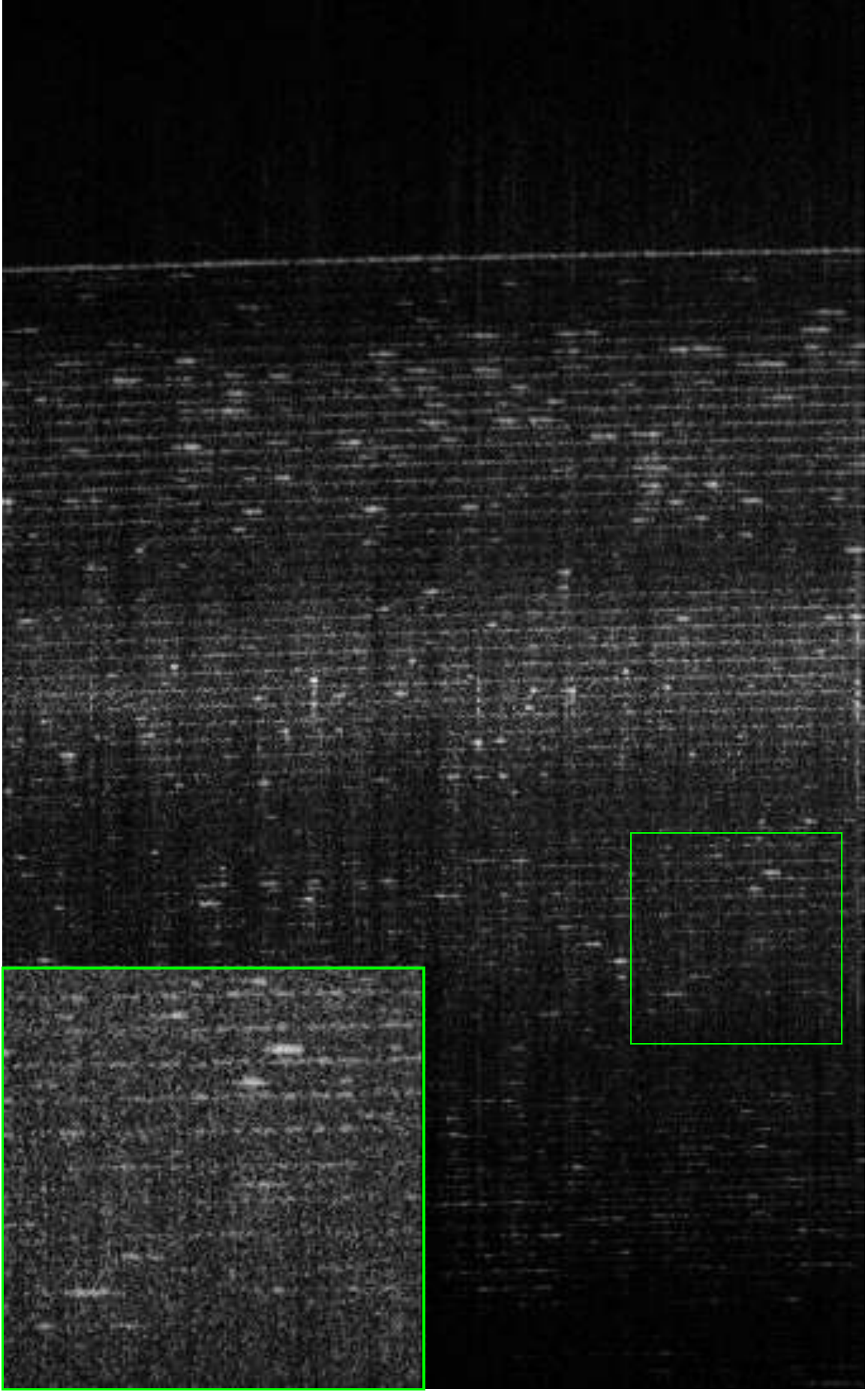}
			\caption{tape DEFR}
		\end{subfigure}
		\hfil
		\begin{subfigure}[b]{0.32\textwidth}
			\includegraphics[width=\textwidth]{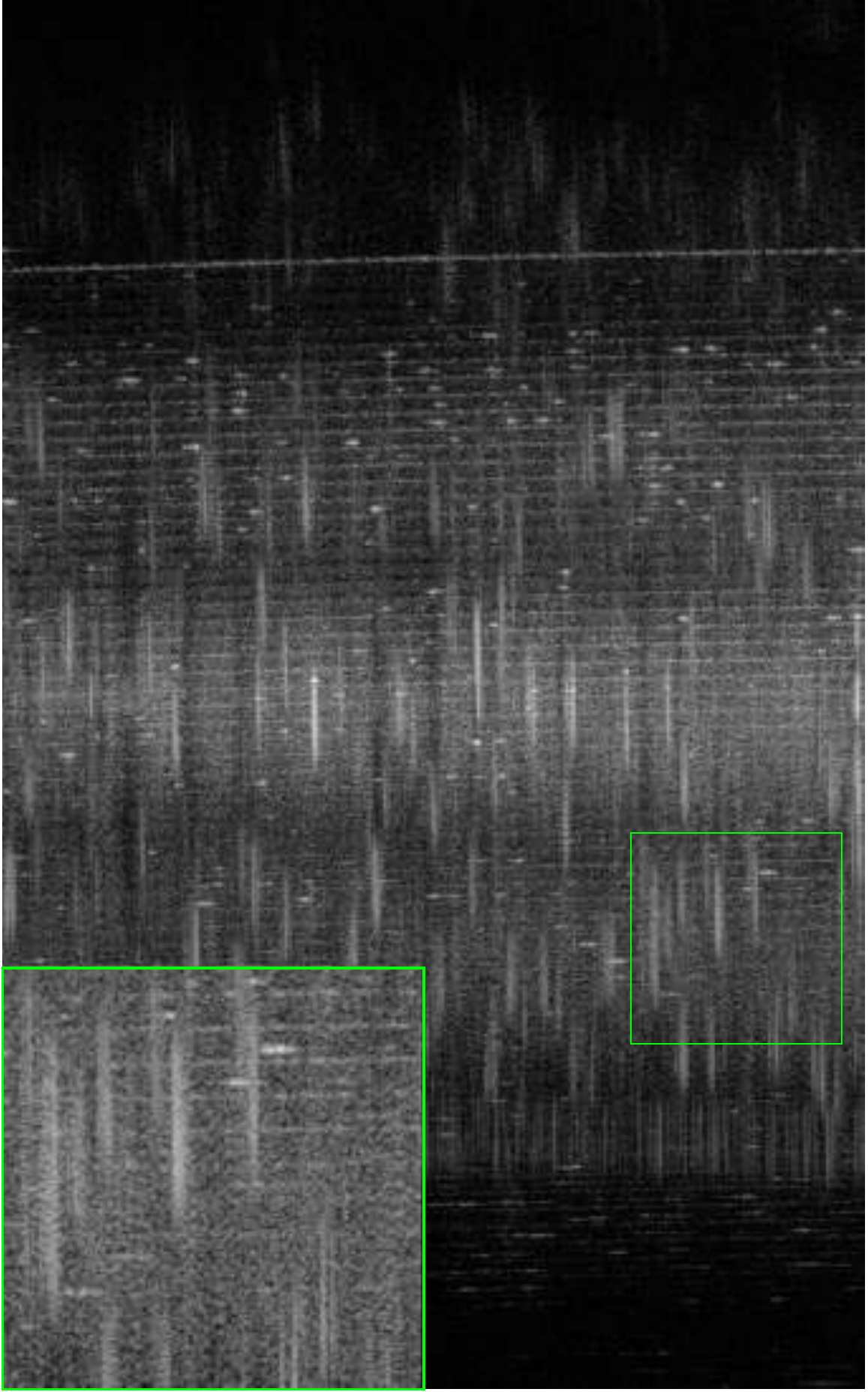}
			\caption{tape ISAM}
		\end{subfigure}
		
		\begin{subfigure}[b]{0.32\textwidth}
			\includegraphics[width=\textwidth]{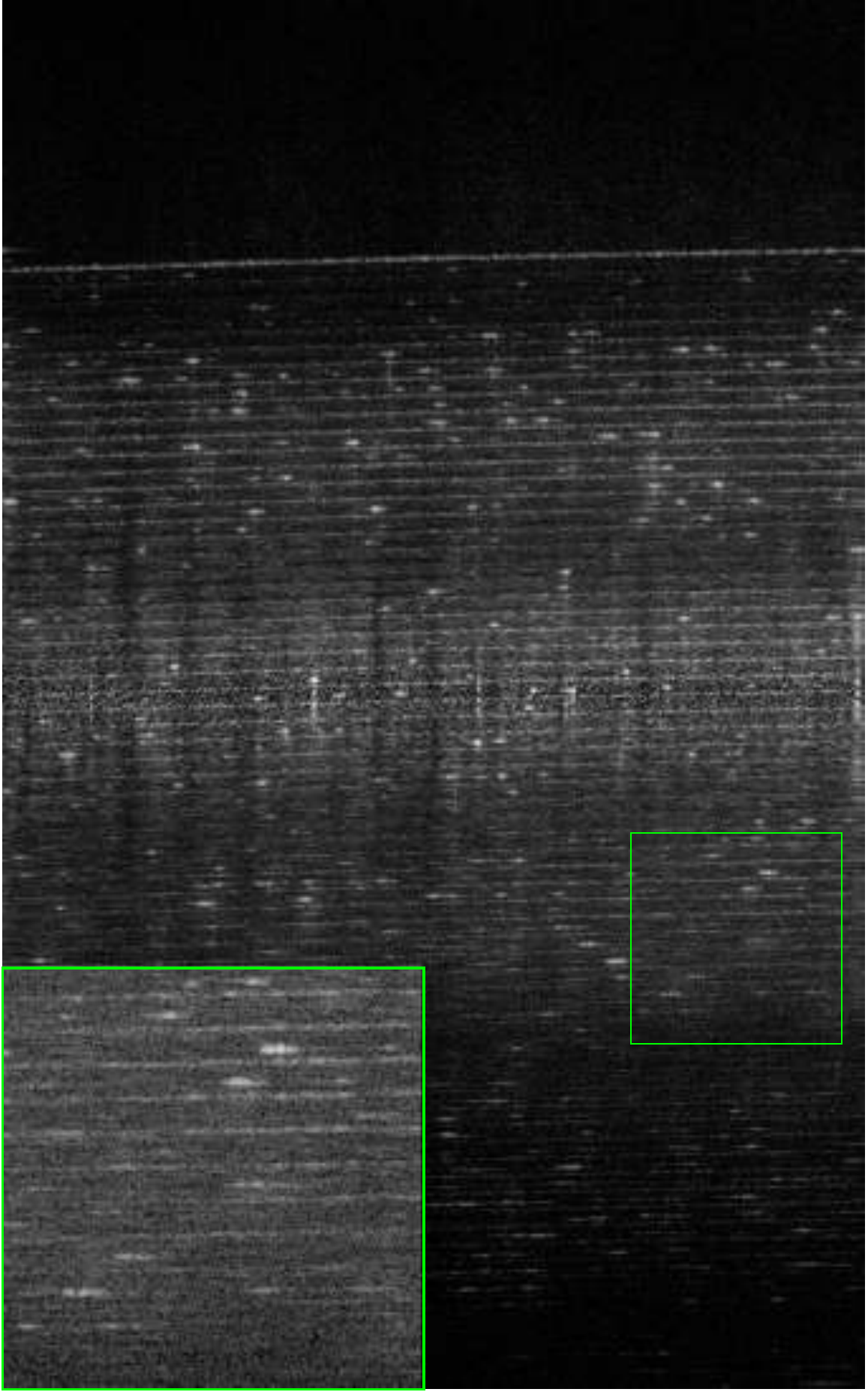}
			\caption{tape DEFR+ISAM}
		\end{subfigure}
		\hfil
		\begin{subfigure}[b]{0.32\textwidth}
			\includegraphics[width=\textwidth]{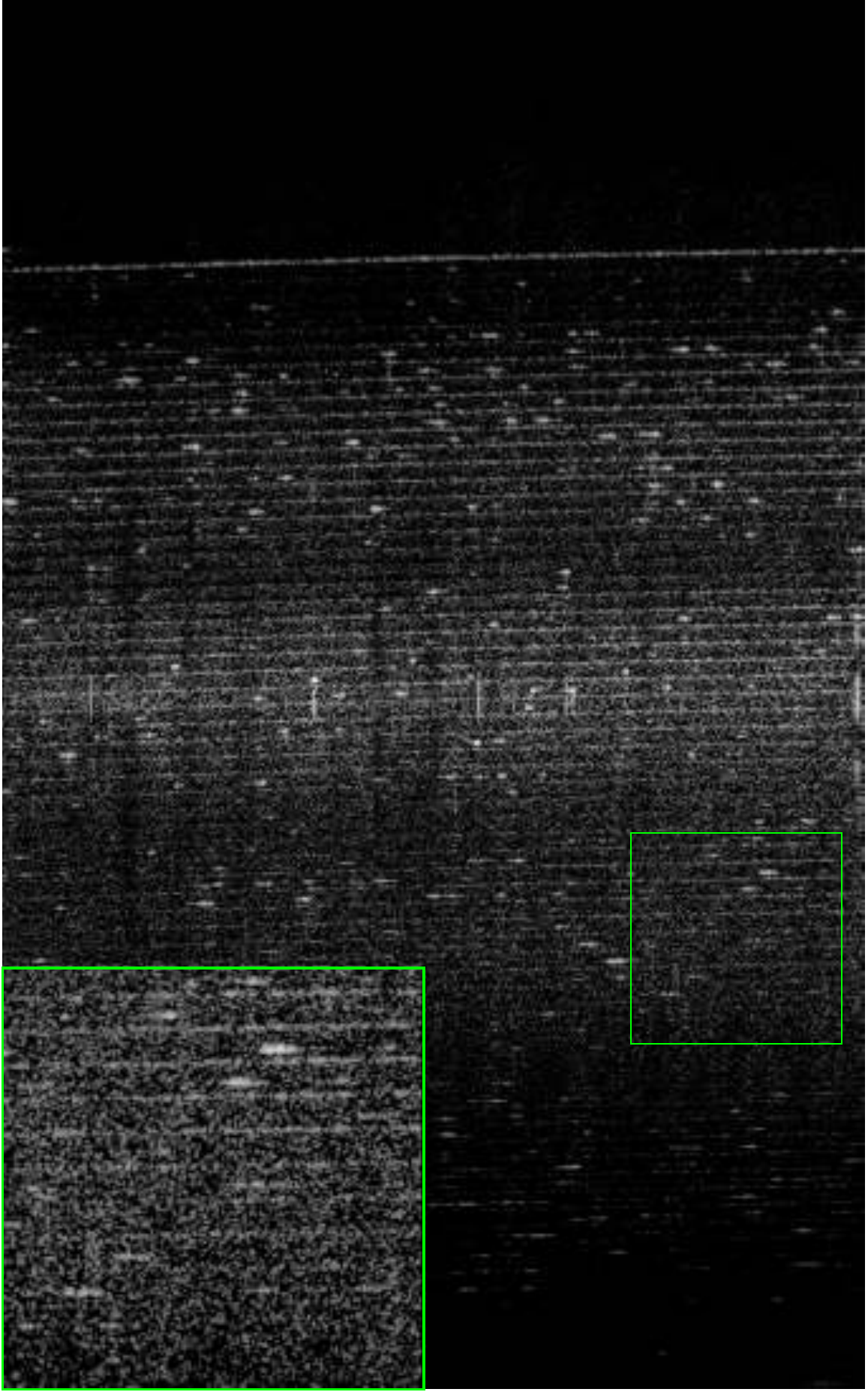}
			\caption{tape MBIR}
		\end{subfigure}
		\hfil
		\begin{subfigure}[b]{0.32\textwidth}
			\includegraphics[width=\textwidth]{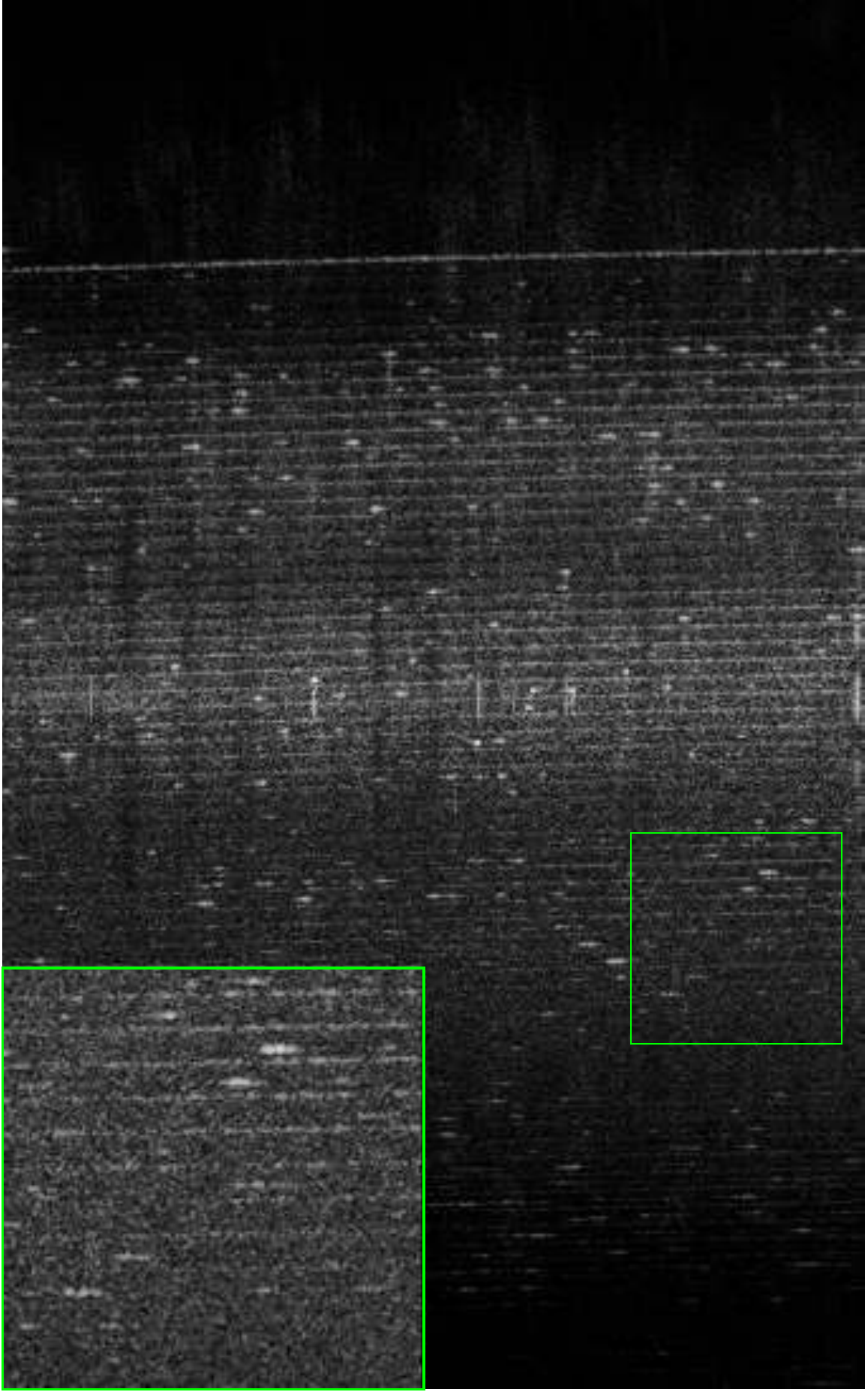}
			\caption{tape MBIR+}
		\end{subfigure}
		\caption{\label{fig:samples_st}Reconstructions of Scotch tape sample with real measurements. The various methods and their implementation are detailed in Sec.~\ref{sec:methods}.}
	\end{figure}
	\begin{figure}[!htbp]
		\centering
		\begin{subfigure}[b]{0.32\textwidth}
			\includegraphics[width=\textwidth]{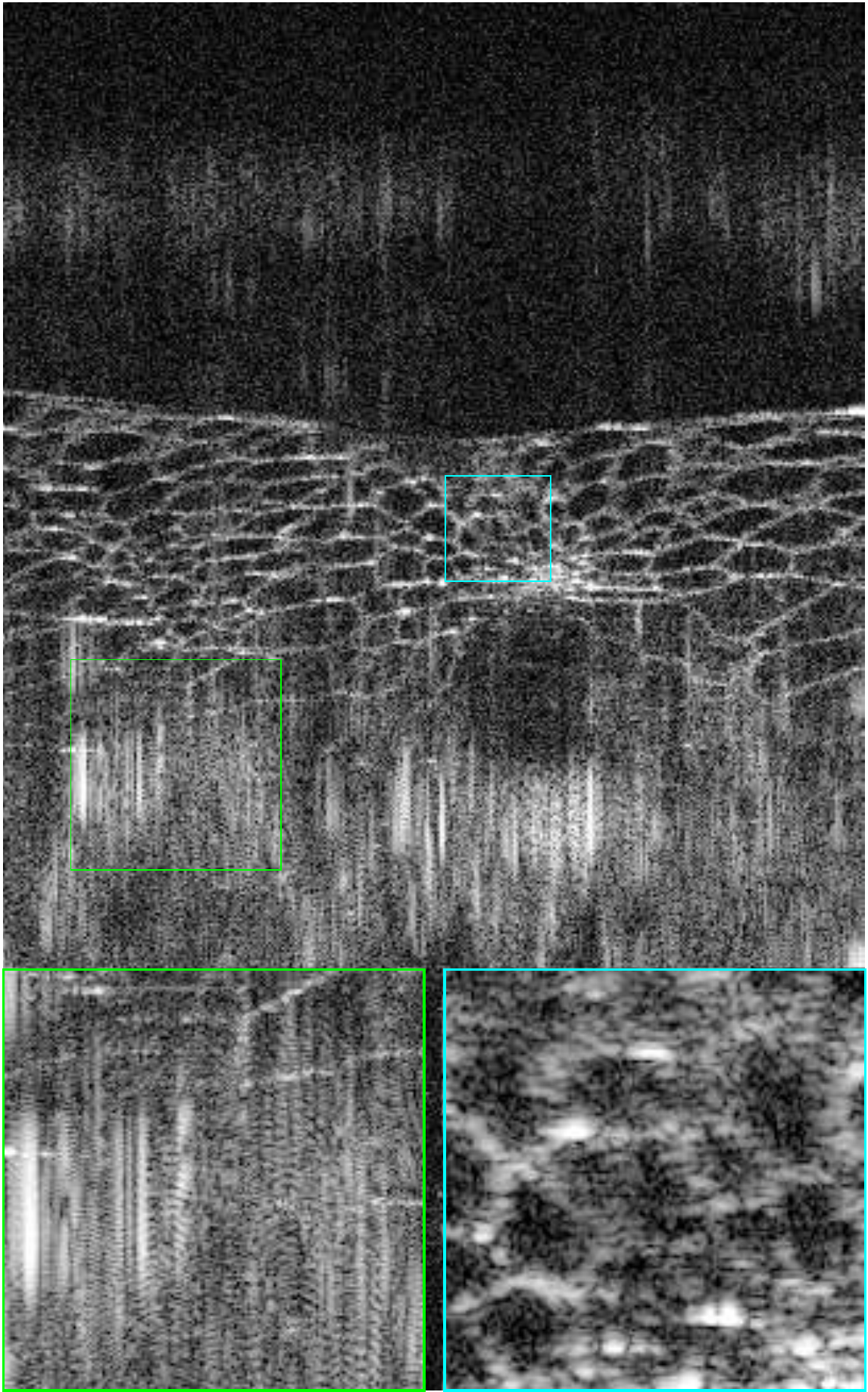}
			\caption{cucumber direct}
		\end{subfigure}
		\hfil
		\begin{subfigure}[b]{0.32\textwidth}
			\includegraphics[width=\textwidth]{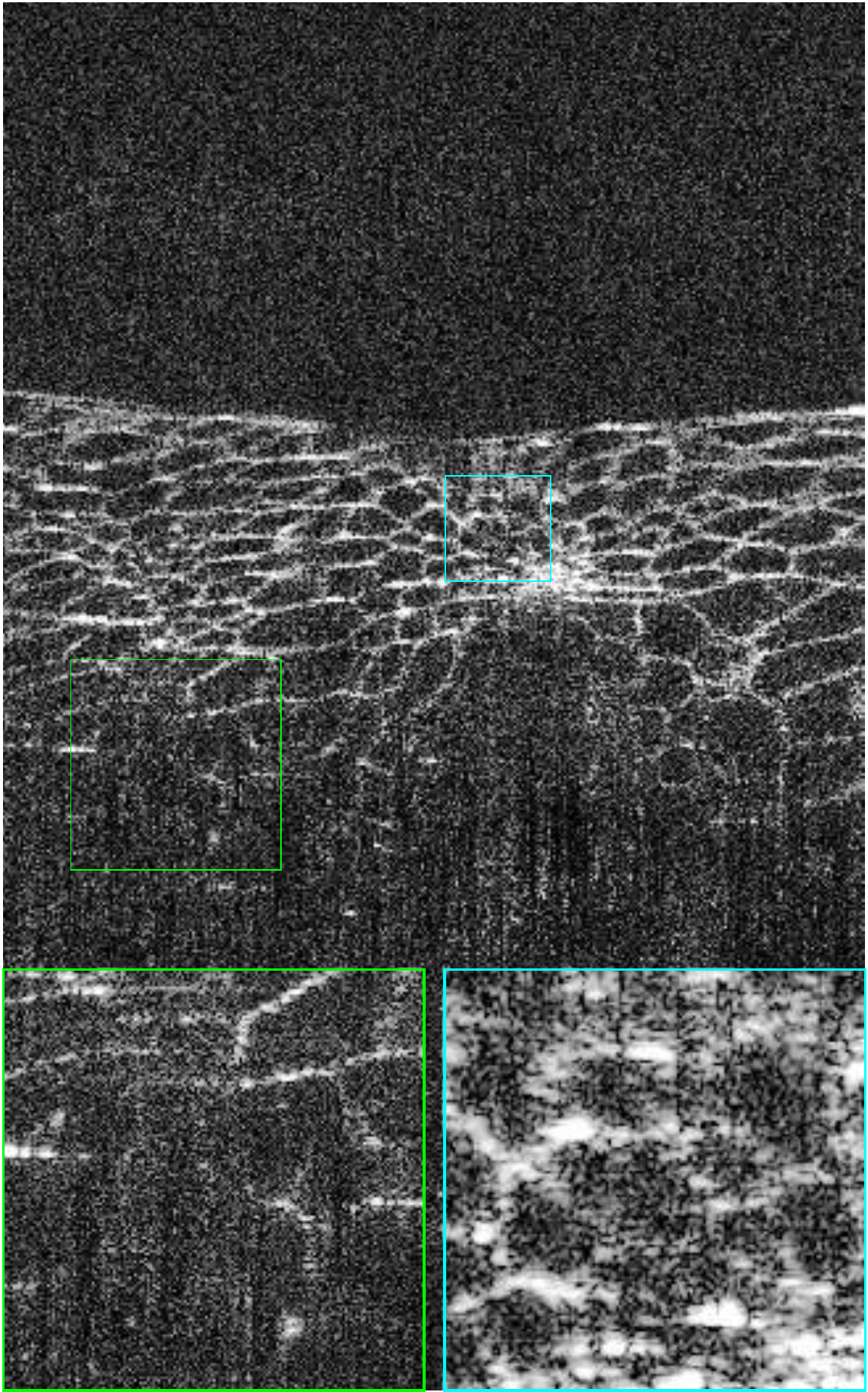}
			\caption{cucumber DEFR}
		\end{subfigure}
		\hfil
		\begin{subfigure}[b]{0.32\textwidth}
			\includegraphics[width=\textwidth]{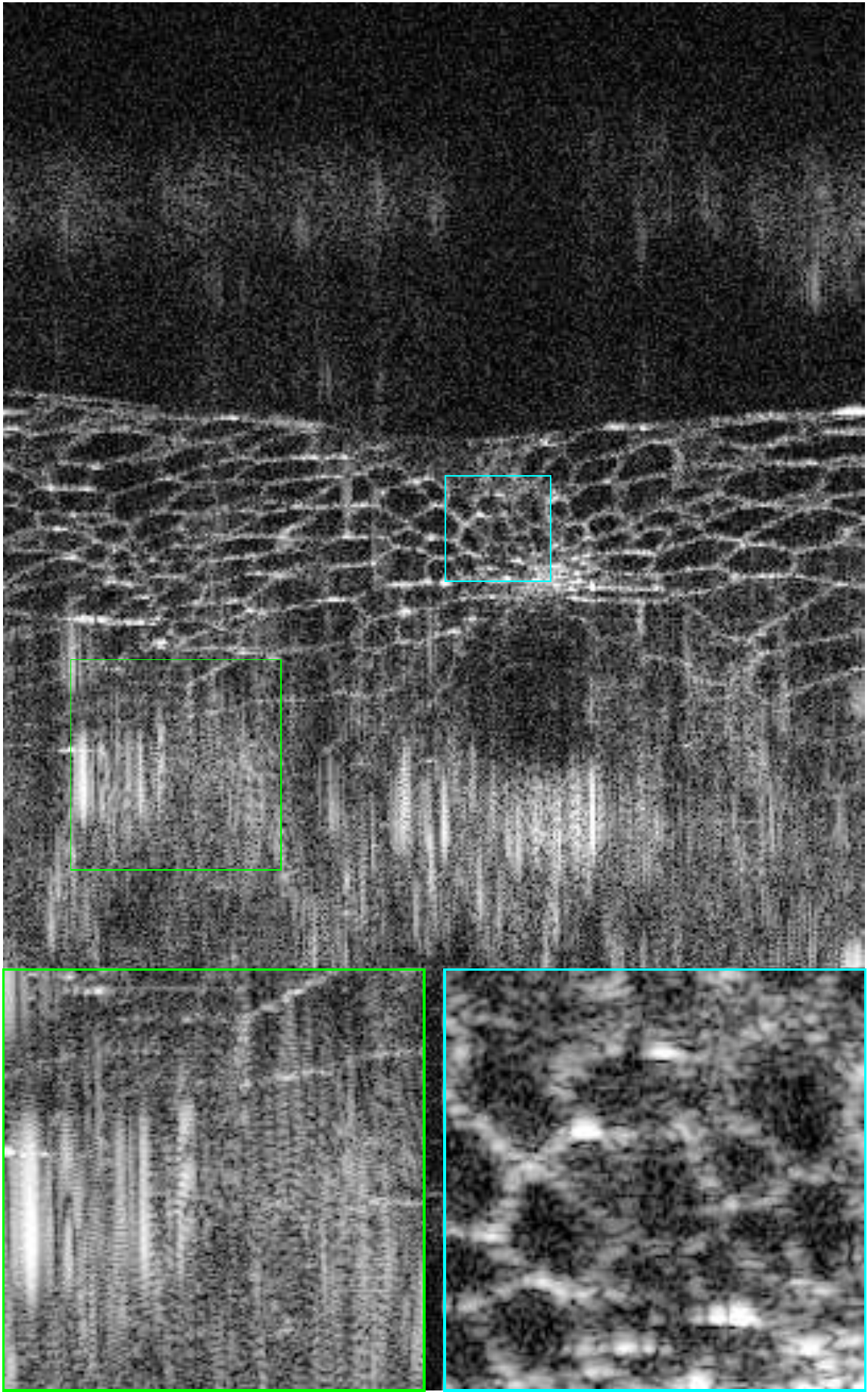}
			\caption{cucumber ISAM}
		\end{subfigure}
		\begin{subfigure}[b]{0.32\textwidth}
			\includegraphics[width=\textwidth]{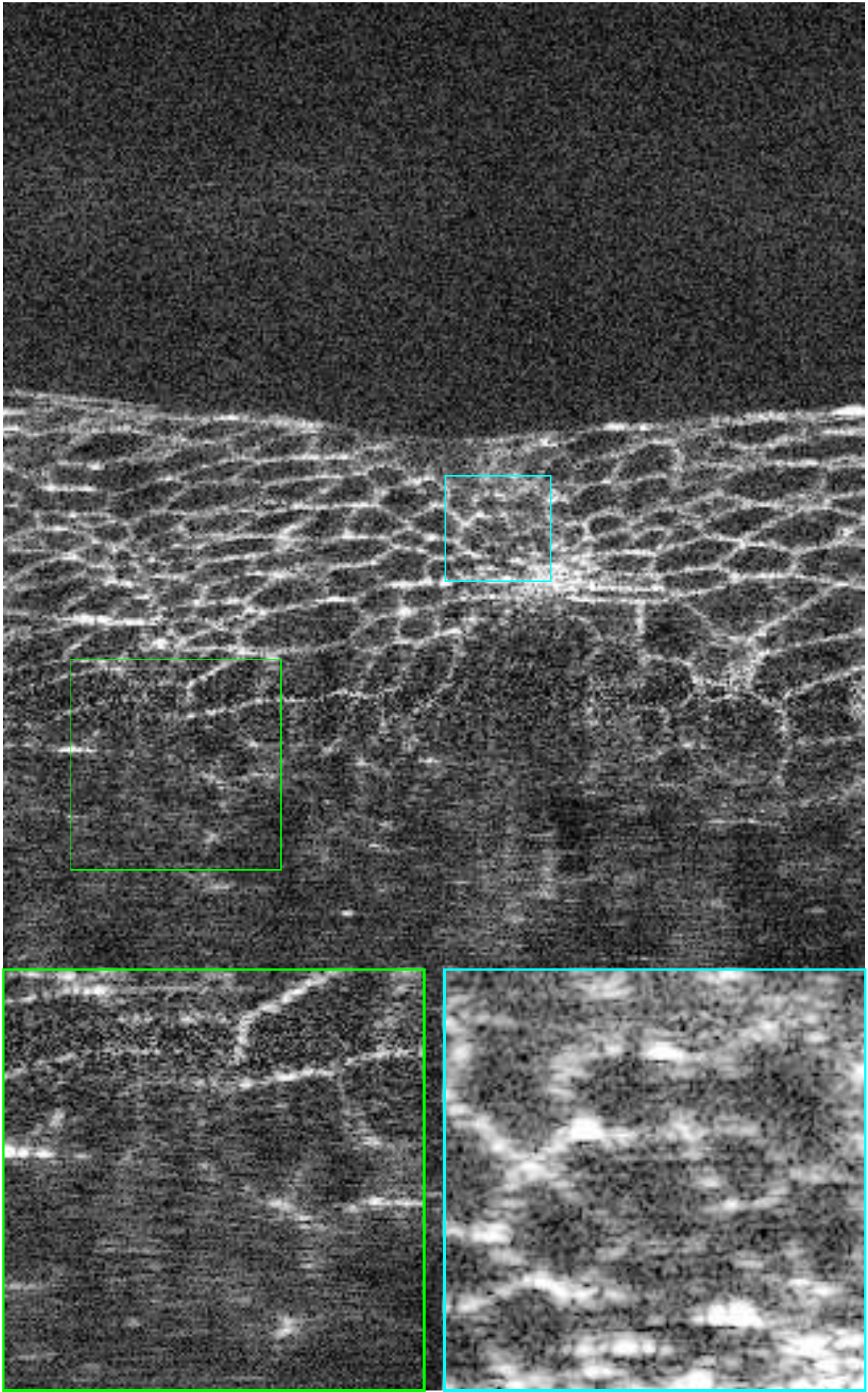}
			\caption{cucumber DEFR+ISAM}
		\end{subfigure}
		\hfil
		\begin{subfigure}[b]{0.32\textwidth}
			\includegraphics[width=\textwidth]{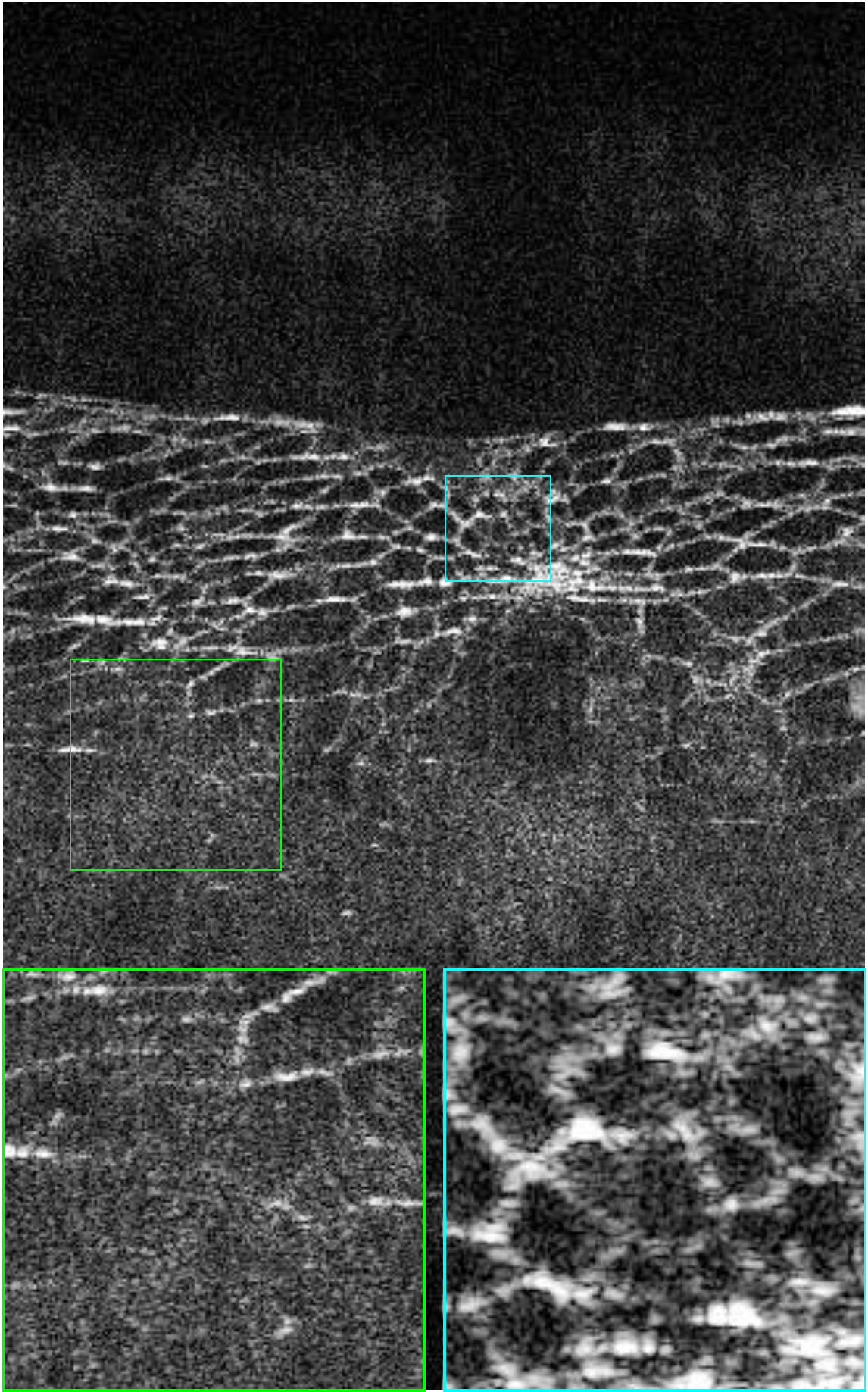}
			\caption{cucumber MBIR}
		\end{subfigure}
		\hfil
		\begin{subfigure}[b]{0.32\textwidth}
			\includegraphics[width=\textwidth]{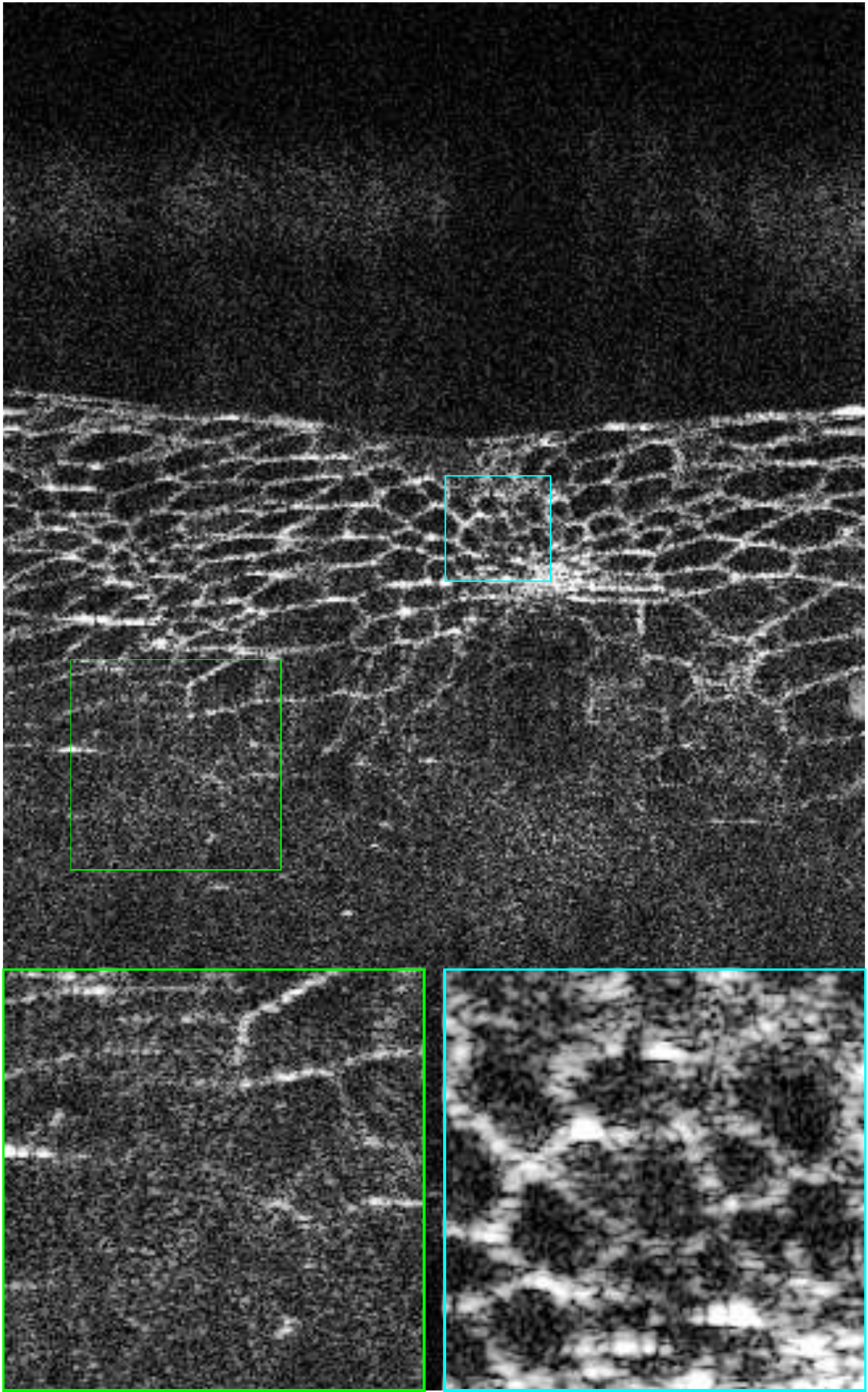}
			\caption{cucumber MBIR+}
		\end{subfigure}
		\caption{\label{fig:samples_cu1}Reconstructions of cucumber with real measurements. The various methods and their implementation are detailed in Sec.~\ref{sec:methods}.}
	\end{figure}
	
	Firstly, we have shown reconstructions from the same beaded gel and cucumber samples as used in Sec.~\ref{sec:quant}. The samples are shown in Fig.~\ref{fig:samples_ti}, Fig.~\ref{fig:samples_st} and Fig.~\ref{fig:samples_cu1} respectively, but with the focal plane positioned at the zero-delay position and taking the fully-sampled spectroscopic data.
	
	%
	
	Several observations can be made from the full-range reconstruction of the beaded gel and tape images in Fig.~\ref{fig:samples_ti} and Fig.~\ref{fig:samples_st} respectively. Similarly to the images in Sec~\ref{sec:quant}, DEFR is effective at suppressing conjugate components, ISAM is effective in refocussing blurred structures away from the focal plane, and DEFR+ISAM or MBIR combine both of these benefits. Unlike the DEFR+ISAM, where there is visible lateral blurring around some of the structures, especially in the inset images, both MBIR and MBIR+ produce sharper looking images. Another feature from DEFR+ISAM, especially of the beaded gel, is an apparent double lobe effect above and below the focal plane. Since this is not present in any other image, it is likely an artifact, but one that does not seem to affect MBIR. We suggest it is an effect of subtracting the conjugate components, which will only be approximated by the implicit DEFR dispersion model, and can hence lead to cumulative errors, especially in areas around the focal plane with greatest intensity. Between the MBIR+ and MBIR, the former does have visibly reduced residual artifacts, whilst preserving all structural information.
	
	From the cucumber tissue images in Fig.~\ref{fig:samples_cu1}, the gain in structural clarity of MBIR+ over DEFR+ISAM can be seen in the inset images. The cell boundaries have greater definition, and some fine structures that are distorted in the other cases, are visible with MBIR. This further confirms the advantage of our proposed method.
	
	In our tested samples, there are no significant autocorrelation artifacts around the zero delay visible, but we suggest these are likely to increase with samples of higher scattering intensity.
	\clearpage
	\section{Conclusions}
	We have demonstrated full-range ISAM through dispersion encoding, and presented a MBIR algorithm for its implementation. While an naive DEFR+ISAM is an alternative way to achieve this, it does not fully exploit the ISAM signal model or multidimensional sparsity, resulting in structural degradation and observed artifacts. In contrast, MBIR produces images of enhanced structural clarity and significantly lower errors. Within this, we have adopted an efficient NUFFT implementation of ISAM, which is numerically stable through many iterations. Ongoing and future work includes reducing the computational time of MBIR, actively modelling autocorrelation artifacts, extending the method to work efficiently with 3D datasets, and to explore compelling biomedical applications.
	
	\section*{Funding}
	UK Engineering and Physical Sciences Research Council (EP/P031250/1); European Research Council (ERC-ADG-2015-694888); Royal Society Wolfson Research Merit Award.
	
	\section*{Acknowledgments}
	The authors sincerely thank Graham Anderson from the University of Edinburgh, for assistance creating the beaded gel phantom. We acknowledge NVIDIA Corporation for kindly donating the Titan Xp GPU. This work has made use of the resources provided by the Edinburgh Compute and Data Facility (ECDF).
	
	\section*{Disclosures}
	The authors declare that there are no conflicts of interest related to this article.
	
	\bibliographystyle{apalike}
	\bibliography{refs}
\end{document}